\documentclass[prb,twocolumn,amsmath,amssymb,superscriptaddress,floatfix,nofootinbib]{revtex4-2}
\usepackage{bm}                     
\usepackage{appendix}
\usepackage[latin1]{inputenc}
\usepackage{dcolumn}      
\usepackage{bbm}
\usepackage{color}
\usepackage{xcolor}
\usepackage{adjustbox}

\usepackage[mathscr]{eucal}
\usepackage{latexsym}
\usepackage{amsbsy} 
\usepackage{graphicx}
\usepackage{epsfig}
\usepackage{subfigure}
\usepackage{wrapfig}
\usepackage{epsf} 
\usepackage[normalem]{ulem}
\usepackage{qcircuit}

\definecolor{darkblue}{HTML}{004D6B}
\definecolor{darkred}{HTML}{8c1515}
\definecolor{darkgreen}{HTML}{006400}
\usepackage{hyperref}
\hypersetup{ colorlinks=true, urlcolor=darkblue, citecolor=darkred,
    linkcolor=darkblue, breaklinks }

\newcommand{\ba}{\begin{array}}
\newcommand{\ea}{\end{array}}
\newcommand{\be}{\begin{equation}}
\newcommand{\ee}{\end{equation}}
\newcommand{\bea}{\begin{eqnarray}}
\newcommand{\eea}{\end{eqnarray}}

\begin{document}

\title{Nonreciprocal charge transport and subharmonic structure
 in voltage-biased Josephson diodes}

\author{A. Zazunov}
\affiliation{Institut f\"ur Theoretische Physik, Heinrich-Heine-Universit\"at, D-40225  D\"usseldorf, Germany}
\author{J. Rech}
\affiliation{Aix Marseille Univ., Universit\'e de Toulon, CNRS, CPT, Marseille, France}
\author{T. Jonckheere}
\affiliation{Aix Marseille Univ., Universit\'e de Toulon, CNRS, CPT, Marseille, France}
\author{B. Gr{\'e}maud}
\affiliation{Aix Marseille Univ., Universit\'e de Toulon, CNRS, CPT, Marseille, France}
\author{T. Martin}
\affiliation{Aix Marseille Univ., Universit\'e de Toulon, CNRS, CPT, Marseille, France}
\author{R. Egger}
\affiliation{Institut f\"ur Theoretische Physik, Heinrich-Heine-Universit\"at, D-40225  D\"usseldorf, Germany}

\date{\today}

\begin{abstract}
We study charge transport  in voltage-biased single-channel junctions involving helical
superconductors with finite Cooper pair momentum. For a Josephson junction, the equilibrium
current-phase relation shows a superconducting diode effect: the critical current depends on
the propagation direction.   We formulate a scattering theory for voltage-biased Josephson diodes
and show that multiple Andreev reflection processes cause a rich subharmonic structure in the DC current-voltage curve at low temperatures and small voltages due to Doppler shifts of the spectral gap. 
In the current-biased case, the diode efficiency has maximal rectification efficiency $\eta_0\approx 0.4$ for this model.  In the voltage-biased case, however, the rectification 
efficiency can reach the ideal value $\eta=1$.
We also discuss charge transport for NS junctions between a normal metal and a helical 
superconductor and comment on related models with spin-orbit interactions and magnetic Zeeman 
fields.
\end{abstract}
\maketitle

\section{Introduction}\label{sec1}

Following the first experimental observations in 2020, the superconducting diode effect (SDE) attracted a huge wave of interest and  
has already been confirmed in several platforms 
\cite{Ando2020,Lyu2021,Bauriedl2022,Baumgartner2022,Pal2022,Lin2022,Wu2022a,Jeon2022,Turini2022,Sundaresh2023,Mazur2023,Anwar2023,Banerjee2023,Ghosh2023,Hou2023a,Costa2023}, see Ref.~\cite{Nadeem2023} for a review.  
In simple terms, the SDE is realized if a superconductor or a Josephson junction sustains a direction-dependent supercurrent, where the equilibrium 
critical supercurrent $I_{c+}>0$ flowing, say, to the right is different
from the critical current $|I_{c-}|$ flowing to the left (with $I_{c-}<0$).  Consider, for instance, the case $|I_{c-}|<I_{c+}$, where this asymmetry implies
that a dissipationless supercurrent $I$ with $|I_{c-}|<|I|<I_{c+}$ can only flow to the right but the corresponding current in the opposite direction must be dissipative.
This equilibrium SDE has been studied by most previous works and can be quantified in terms of an
SDE efficiency parameter $0\le \eta_0\le 1$ defined by
\begin{equation}\label{eta0}
\eta_0 =\frac{I_{c+}-|I_{c-}|}{I_{c+}+|I_{c-}|},
\end{equation} 
 where $\eta_0=1$ refers to the ideal SDE.  
While the SDE has been predicted many years ago \cite{Edelstein1996,Hu2007,Reynoso2008,Zazunov2009}, recent theoretical work
has clarified how this physics appears in different platforms or device geometries
\cite{Misaki2021,Ilic2022,He2022,Zhang2022,Davydova2022,FuErratum,Kokkeler2022,Daido2022a,Daido2022b,Tanaka2022,Zinkl2022,Yuan2022,Legg2022,Halterman2022,Cheng2023,Ikeda2023,He2023a,Lu2023,Fu2023,Wang2023a,Yuan2023b,Legg2023,Nakamura2023,Picoli2023}.
However,  the detailed microscopic mechanisms behind the SDE in a given experiment are often not well understood.  
In any case, given the rapid progress in this field, it seems likely that  superconducting diodes will soon find technological applications \cite{Nadeem2023}, e.g., as efficient
sensors for magnetic textures \cite{Sinner2023}.

While the SDE is possible in junction-free bulk superconductors \cite{Edelstein1996,Ando2020,Bauriedl2022,Nadeem2023}, we here study nonequilibrium charge transport for
 the case of a single Josephson junction exhibiting the SDE in equilibrium.  In contrast to the resulting ``intrinsic'' SDE, the ``extrinsic'' Josephson diodes arise as a consequence of having more complicated device geometries 
with several junctions \cite{Souto2022,Fominov2022,Paolucci2023,Gupta2023,Ciaccia2023,Zhang2023expA}.
Such setups
contain more tuning parameters and useful applications may thus be easier to identify.  
For the (perhaps more fundamental) intrinsic SDE, one usually requires that time-reversal symmetry and inversion symmetry have to be broken and, in addition, the 
equilibrium current-phase relation (CPR) must be non-sinusoidal \cite{Reynoso2008,Zazunov2009,Bauriedl2022,Baumgartner2022}. The latter requirement typically implies 
that the junction should not be in the deep tunneling limit. For the single-channel case studied below, this means that the transmission probability $0<{\cal T}\le 1$ of the junction should be as large as possible. In practice, the SDE is   accompanied by the anomalous Josephson effect \cite{Buzdin2008,Reynoso2008,Zazunov2009,Reynoso2012,Brunetti2013,Dolcini2015,Campagnano2015,Szombati2016,Qin2017,Minutillo2018,Alidoust2021},  that is, the supercurrent is finite at vanishing phase difference. 

It is well known that magnetochiral effects \cite{Rikken2001,Tokura2018,Morimoto2018} can be responsible 
for the SDE in noncentrosymmetric superconductors \cite{Edelstein1996}. 
An alternative mechanism arises if the superconductor has finite Cooper pair
(CP) momentum $2q\ne 0$ (``helical superconductivity'') \cite{Davydova2022,Yuan2022,Pal2022,Lin2022,Banerjee2023}.
Such type of spin-singlet pairing has been reported, e.g., for the orbital Fulde-Ferrell-Larkin-Ovchinnikov state in the multilayer Ising superconductor 2H-NbSe$_2$ \cite{Wan2023}.   
We here mainly consider charge transport through a short single-channel weak link 
connecting two helical superconductors, but we briefly comment on a different model based on the interplay of spin-orbit coupling and a magnetic Zeeman field, see App.~\ref{appA}. Such models can yield a magnetochiral SDE. We demonstrate that
scattering theory then yields the exact $I$-$V$ curve for this model, where
known results \cite{Bratus1995,Averin1995,Zazunov2006} are recovered for $q=0$.      In addition, we also address charge transport in  NS junctions between a normal metal and a helical superconductor.

For the Josephson diode case, for simplicity, we assume identical pairing gap $\Delta$ and CP momentum $2q$ on both sides, where the CPR exhibits the SDE \cite{Davydova2022}.  Key parameters are the transparency ${\cal T}$ of the weak link and the dimensionless CP momentum parameter $q\xi$, where the superconducting coherence
length is  $\xi=\hbar v_F/\Delta$ with the Fermi velocity $v_F$.  Without loss of generality,
we assume $q\ge 0$.  In addition, in order to retain a spectral gap in the superconductor, we focus on the parameter regime $q\xi<1$.  Since the resulting voltage-biased Josephson diode model is relatively simple,  the exact computation of the DC current-voltage ($I$-$V$) curve from scattering theory taking into account multiple Andreev reflection (MAR) processes \cite{Klapwijk1982,Bratus1995,Averin1995,Cuevas1996} is possible.  
Throughout we focus on the most interesting low-temperature regime where MAR can enable efficient charge transfer across the junction, 
especially for voltages in the subgap regime $e|V|< 2\Delta$. 
 We focus on junctions with a single (or a few uncoupled) 
channels, where the impedance is of order $h/e^2$ and thus much larger than the typical
impedance of the external circuit. We then do not have to account for
the self-consistent dynamics of the phase difference and voltage across the junction. 
We note that previous  work on nonequilibrium charge transport in Josephson diodes has studied weakly damped  junctions \cite{Misaki2021,Fominov2022,Trahms2023,Steiner2023} and externally driven junctions \cite{Paaske2023}, but MAR effects have not been addressed to our knowledge.

Under nonequilibrium conditions corresponding to a constant bias voltage $V$ across the junction, charge transport rectification appears if $I(-V)\ne -I(V)$. In analogy to the SDE
efficiency $\eta_0$ in Eq.~\eqref{eta0}, we quantify the finite-voltage rectification efficiency by the
dimensionless parameter
\begin{equation}\label{efficiency}
    \eta(V) = \frac{I(V)+I(-V)}{I(V)-I(-V)},
\end{equation}
which vanishes for $q=0$.    
Our theory predicts a characteristic voltage-dependent 
rectification pattern, where $\eta(V)$ is especially large in the subgap regime. 
For the conventional case without SDE, MAR causes a subharmonic structure,
i.e., singular features in the nonlinear conductance for $eV=2\Delta/n$ with integer $n$ \cite{Klapwijk1982,Bratus1995,Averin1995}.   
For Josephson diodes, we predict an even richer subharmonic structure which determines
the rectification characteristics and might provide precious information about 
the microscopic mechanisms generating the SDE. 
Our central finding is that the efficiency $\eta(V)$ can approach the ideal limit of full rectification with $\eta=1$
at low voltages, even though $\eta_0\alt 0.4$ for the SDE efficiency in equilibrium for the model 
considered below. The importance of MAR processes 
for rectification is related to the fact that higher harmonics of the CPR are 
needed for the SDE. Indeed, equilibrium Andreev states are the result of resonant MAR processes.  
A finite voltage breaks up the resonant MAR loop (see below) 
and effectively opens the way to high anharmonicity with many harmonics.  
In contrast to the current-biased case \cite{Davydova2022,FuErratum}, 
we find that the voltage-biased junction has a different optimal working point
and allows for ideal rectification.  We note that this model has been successfully used
to explain experimental results for current-biased Josephson diodes \cite{Pal2022} reporting a maximum
efficiency $\eta_0\approx 0.4$. 

The structure of the remainder of this article is as follows. 
 In Sec.~\ref{sec2}, we describe the model  and the corresponding eigenstates for a weak link between two  helical superconductors with finite CP momentum.  The
  equilibrium CPR and the resulting SDE efficiency $\eta_0$ are 
 discussed in Sec.~\ref{sec3}. In Appendix \ref{appA}, we comment on the SDE efficiency for a different Josephson diode model,
where one has conventional superconductors with $q=0$ but the weak link corresponds 
to a quantum dot with spin-orbit coupling and a magnetic Zeeman field \cite{DellAnna2007,Zazunov2009,Brunetti2013}.  Effectively, the SDE is then caused 
by magnetochiral anisotropy. Subsequently, charge transport through NS junctions involving a helical superconductor is addressed in Sec.~\ref{sec4}.
 We then present the MAR scattering theory for a voltage-biased Josephson junction in Sec.~\ref{sec5}.  The rectification properties out of equilibrium are then discussed in Sec.~\ref{sec6}. The paper concludes with a summary and an outlook
in Sec.~\ref{sec7}.  We often put $\hbar=1$.

%%%%%%%%%%%%%%%%%%%%%%%%%%%%%%%%%%%%%%%%%%%%%%%%%%%%%%%%%%%%%%%%%%%%%%%%%%%%%%%%%%%%%%%%%%
\section{Helical superconductor junction}\label{sec2}

In Sec.~\ref{sec2a}, we describe the model for a short single-channel Josephson diode involving helical superconductors \cite{Davydova2022}. 
The spectrum and the eigenstates of the corresponding Bogoliubov-de Gennes (BdG) problem are then summarized in Sec.~\ref{sec2b}.

\subsection{Model}\label{sec2a}

Let us consider a short Josephson junction with a weak link connecting two $s$-wave BCS superconducting banks with identical pairing gap $\Delta$ 
and coherence length $\xi$.
The superconductors are described in the quasiclassical  Andreev approximation valid for $k_F\xi\gg 1$  with Fermi momentum $k_F$. By linearizing the band dispersion around $\pm k_F$,
the full Nambu spinor $\psi(x,t)$ is written in terms of Nambu spinor envelopes
 $\psi_{\alpha=\pm}(x,t)$ for states with momenta near $\pm k_F$,
\begin{eqnarray} \label{nambu}
 \psi(x,t)&=&\sum_\pm e^{\pm ik_Fx} \psi_\pm(x,t),\\ \nonumber
 \psi_{\alpha=\pm}(x,t)&=& \sum_k e^{ikx} \left(\begin{array}{c} \psi_{\alpha k_F+k,\uparrow}(t)\\ 
 \psi_{-(\alpha k_F+k),\downarrow}^\dagger (t)\end{array}\right), \quad |k|\ll k_F.
\end{eqnarray}
For instance, $\psi_+$ refers to right-moving electron-like and left-moving hole-like quasiparticles.   
With indices $s_{L}=1$ and $s_R=-1$, and using
 Pauli matrices $\tau_{x,y,z}$ (and the identity $\tau_0$) in Nambu space,
the Hamiltonian $H_{L/R}$ for the left ($x<0$) and right ($x>0$) superconductor, respectively, has the effectively one-dimensional form 
\begin{eqnarray} \nonumber
  &&  H_{j=L/R}(t) = \sum_{\pm} \int_{s_jx<0} dx\, \psi^\dagger_\pm(x,t) \bigl ( 
\mp i  v_F  \tau_z \partial_x + \\ &&\qquad + e V_j(t) \tau_z + \Delta  \tau_x e^{i \tau_z [-2qx+\phi_j(t)]}
\bigr)\psi^{}_\pm(x,t), \label{HLR}
\end{eqnarray}
where the voltages  $V_{j}(t)$ and superconducting phases $\phi_j(t)$ are linked by the Josephson relation, $eV_j=\dot \phi_j/2$. 
The gauge-invariant phase difference across the contact is $\varphi(t)=\phi_L(t)-\phi_R(t)$. 
A finite CP momentum  $2q\ne 0$ breaks time-reversal and inversion symmetries and can generate the SDE.  
Microscopic mechanisms generating $q\ne 0$ have recently been discussed in Refs.~\cite{Davydova2022,Yuan2022,Levichev2023}.
For the corresponding helical superconductor, the pairing order parameter oscillates in space, $\Delta_j(x)=\Delta e^{-i\phi_j+2iqx}$,
where we assume $0\le q\xi<1$.

Modeling the weak link at $x=0$ as a single-channel normal-conducting constriction of short length $\ll \xi$
and arbitrary transmission probability ${\cal T}$, the quasiclassical envelopes on both sides of the contact
 $(x=0^\pm)$ are matched by a transfer matrix \cite{Zazunov2005,Nazarov2009,Zazunov2014,Ackermann2023},
\begin{equation}\label{BC1}
\Psi(0^-,t) = \frac{1}{\sqrt{\cal T}}(\sigma_0+r\sigma_x)\tau_0 \Psi(0^+,t),
\end{equation}
with the reflection amplitude $r=\sqrt{1-{\cal T}}$.  The
Pauli matrices $\sigma_{x,y,z}$ (identity $\sigma_0$) 
and the bispinor $\Psi(x,t)=(\psi_+,\psi_-)^T$ act in right-left mover space, with the Nambu spinors $\psi_\pm$ in Eq.~\eqref{nambu}. 

In this section, we consider the equilibrium case ($V_j=0$) 
 and choose $\phi_j(t)=s_j\varphi/2$.
(However, our formalism directly carries over to the finite-$V$ case.)
We then perform a unitary transformation to the comoving frame,
\begin{equation}\label{Ugauge}
\Psi(x) \to U(x) \Psi(x),\qquad U= e^{i\sigma_0\tau_z [-qx+\phi_j/2]},
\end{equation}
where the transformed Hamiltonian is  
$H=\sum_j H_j=\int dx \Psi^\dagger {\cal H}_{\rm BdG} \Psi$ with the BdG Hamiltonian (for $x\ne 0$) 
\begin{equation}\label{BdG1}
    {\cal H}_{\rm BdG}= -iv_F \sigma_z\tau_z \partial_x+ v_Fq \sigma_z\tau_0 + \Delta\sigma_0\tau_x.
\end{equation} 
The phase dependence now only appears in the transformed matching condition, 
\begin{equation}\label{BC2}
\Psi(0^-) = \frac{1}{\sqrt{\cal T}} (\sigma_0+r\sigma_x)   e^{i\tau_z \varphi/2}\Psi(0^+).
\end{equation}
BdG eigenstates $\Psi_\nu(x)$ for the respective eigenenergy $E_\nu$ then follow by solving the
BdG problem
\begin{equation}\label{BdG}
    {\cal H}_{\rm BdG} \Psi_\nu(x) =  E_\nu  \Psi_\nu(x)
\end{equation}
with  the matching condition \eqref{BC2} at $x=0$. In what follows, it is useful to define
the Doppler shifted energy $E_\alpha$ for $\alpha$-movers with energy $E$,
\begin{equation}\label{Eal}
E_{\alpha=\pm}= E -\alpha v_Fq,
\end{equation}
which should not be confused with the BdG eigenenergies $E_\nu$.

\subsection{Spectrum and eigenstates}\label{sec2b}

We now show that solutions of the BdG problem \eqref{BdG} include Andreev bound states localized near the junction ($x=0$),
 propagating scattering states corresponding to continuum quasiparticles, and solutions of
mixed character which propagate along one direction but are evanescent in the opposite one. 

\begin{figure}
    \centering
    \includegraphics[width=0.49\textwidth]{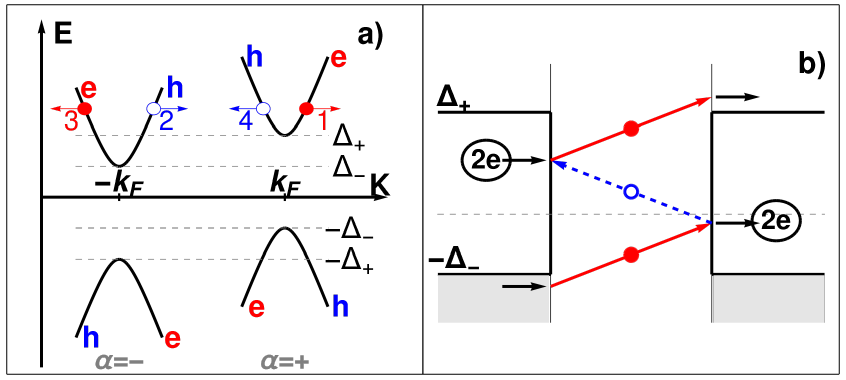}
    \caption{ Schematic illustration of the quasiparticle dispersion and the MAR ladder.  Panel (a) shows
    the dispersion \eqref{bulk} for electron ($e$) and hole ($h$) like states with chirality $\alpha=\pm$.  The four possible types of incident states ($s=1,2,3,4$) are marked by filled (open) circles for electron (hole) states.  Arrows indicate the propagation direction.  Panel (b) shows the MAR ladder picture for the example
    of one MAR trajectory in the ballistic limit. Here an electron like state with incident energy $E$ just
    below $-\Delta_-$ impinges on the junction from the left side, corresponding to $s=1$.  
    By traversing the normal weak link region as right-moving electron (red arrow), it 
    gains the energy $eV$ and is Andreev reflected at the right (NS) interface as left-moving hole (blue dashed arrow).  After gaining another $eV$ quantum, the hole is Andreev reflected at the left (SN) interface.
    The right-moving electron has energy $E_3=E+3eV$ after traversing the normal region and 
    enters the right superconductor.  If $E_3$ is just above $\Delta_+$, a MAR feature corresponding 
    to $eV=2\Delta/3$ will result. 
      }
    \label{fig1}
\end{figure}

Let us start with the effectively junction-free case of a ballistic junction (${\cal T}=1$) at vanishing 
phase difference ($\varphi=0$). In this ``bulk'' case, Eq.~\eqref{BC2} is trivially fulfilled by continuous spinor wave functions.
The resulting BdG eigenstates $\Psi_{\alpha}(x,E)$ are labeled by the conserved chirality $\sigma_z=\alpha=\pm$ and the conserved energy $E$, where
 Eqs.~\eqref{BdG} and \eqref{Eal} imply
\begin{equation}\label{scatt}
  \left(  -i\alpha v_F\tau_z\partial_x+\Delta \tau_x -E_\alpha\tau_0\right) \Psi_{\alpha}(x,E)=0.
\end{equation}
The resulting bulk dispersion for $\alpha$-movers of electron ($e$) or hole ($h$) type 
is given by
\begin{equation}\label{bulk}
    E_{e/h}(k)=  \alpha v_Fq \pm \sqrt{(v_Fk)^2+\Delta^2},
\end{equation}
see Fig.~\ref{fig1}(a) for an illustration. 
For $q\xi<1$, the positive continuum threshold energies (spectral gaps) are therefore given by
\begin{equation}\label{Dpm} 
\Delta_\pm=\Delta\pm v_F q.
\end{equation}
They differ from $\Delta$ because of the Doppler shift
 generated by the CP momentum $2q$.  

We then return to the generic case with ${\cal T}<1$ and/or $\varphi\ne 0$.  The above discussion indicates that 
Andreev bound states can only exist at subgap energies with respect to both spectral gaps ($|E|<\Delta_-$).
While for $|E|>\Delta_+$, continuum quasiparticles can freely propagate to the right and to the left side,
we have mixed-character states for $\Delta_-<|E|<\Delta_+$.  By means of analytic continuation, we can 
describe all three spectral regions in a unified manner (including the finite-voltage case considered later) as follows. For arbitrary energy $E$,
we first introduce the Andreev reflection amplitude $\rho(E)$ for an NS interface,
\begin{equation}\label{tildegamma} 
\rho(E)=\sigma_E  e^{-\tilde\gamma_E} = \left\{\begin{array}{cc}  
\sigma_E \frac{|E|- \sqrt{E^2-\Delta^2}}{\Delta}, & |E|\ge \Delta,\\ & \\
\frac{E-i \sqrt{\Delta^2-E^2}}{\Delta} \equiv e^{-i\gamma_E}, & |E| <\Delta,
\end{array} \right. 
\end{equation}
with $\sigma_E={\rm sgn} (E)$ and  $\gamma_E=\cos^{-1}(E/\Delta)\in (0,\pi)$ for subgap energies with $|E|<\Delta$.  
Note that $\tilde \gamma_E$ is real-valued for $|E|\ge \Delta$ but purely imaginary for subgap energies.

For the conventional case $q=0$, Nambu spinors $\psi_{e/h}(E)$ for incoming states of electron or hole type, with energy $|E|\ge \Delta$, are 
then given by  
\begin{eqnarray}\nonumber
    \psi_e(E)&=&\tau_x \psi_h(E) = \frac{\lambda(E)}{\sqrt2} \left( \begin{array}{c} 1\\ \rho(E) \end{array}\right),\\ 
    \label{psidef}   \lambda(E) &=& \sqrt{\frac{2}{1+\rho^2(E)}}. 
\end{eqnarray}
These states satisfy the standard normalization condition $\psi_{e/h}^\dagger(E)\cdot \psi^{}_{e/h}(E) = 1$.  
For the scattering state construction used below, it is very convenient to also introduce
  ``partially normalized'' Nambu spinors $\tilde \psi_{e/h}(E)$ for  outgoing states, 
 \begin{equation}\label{tildepsi}
    \tilde\psi_e(E)=\tau_x \tilde\psi_h(E) = \frac{1}{\sqrt2} \left( \begin{array}{c} 1\\ \rho(E) \end{array}\right), 
\end{equation}
which are defined for all possible energies $E$. For subgap energies, these states again satisfy the standard
normalization condition  because then $\rho(E)$ is just a complex phase factor.  Otherwise
the unconventional normalization condition
\begin{equation}\label{normal1}
\left.\tilde \psi_{e/h}^\dagger(E)\cdot \tilde\psi_{e/h}(E) \right|_{|E|\ge \Delta}=\frac{1}{\lambda^2(E)} 
\end{equation}  
has to be taken into account.

For CP momentum $2q\ne 0$, we then express the corresponding Nambu spinors $\psi_{\alpha,e/h}(E)$ 
and $\tilde \psi_{\alpha,e/h}(E)$ for states with momenta near $\alpha k_F$ (with $\alpha=\pm$)
by incorporating the Doppler shift $E\to E_\alpha$, see Eq.~\eqref{Eal}, in the above spinors, see Eqs.~\eqref{psidef} and \eqref{tildepsi},
\begin{equation}\label{tildepsi1}
    \psi_{\alpha,e/h}(E)= \psi_{e/h}(E_\alpha) ,\quad \tilde \psi_{\alpha,e/h}(E)= \tilde\psi_{e/h}(E_\alpha).
\end{equation}
These states solve the BdG equation for $x\ne 0$, 
\begin{equation}\label{eom}
    (\pm v_F k_\alpha(E) \tau_z +\Delta\tau_x-E_\alpha\tau_0)\psi_{\alpha,e/h}(E) = 0,
\end{equation}
and similarly for $\tilde\psi_{\alpha,e/h}(E)$,
with the $+$ ($-$) sign for electron (hole) type solutions.
For incoming spinors $\psi_{\alpha,e/h}(E)$, we demand $|E_\alpha|\ge \Delta$, 
but no energy restrictions apply to outgoing spinors $\tilde\psi_{\alpha,e/h}(E)$. 
For arbitrary $E$, with $\sigma_\alpha(E)={\rm sgn}(E_\alpha)$,
the momentum $k_\alpha(E)$ in Eq.~\eqref{eom} is given by
\begin{equation}\label{kdef}
v_F    k_\alpha(E)=  \left\{   \begin{array}{cc}  \sigma_\alpha(E) \sqrt{E_\alpha^2-\Delta^2}, & |E_\alpha|\ge \Delta,\\ 
    i\sqrt{\Delta^2-E_\alpha^2}\equiv iv_F\kappa_\alpha(E), & |E_\alpha| <\Delta. \end{array} \right. 
\end{equation}
This introduces the decay length scale $\kappa_\alpha^{-1}(E)$ for evanescent states with $|E_\alpha|<\Delta$.
 
For given energy $E$, a BdG eigenstate can then be written in terms of incoming and outgoing states, 
\begin{equation} \label{Ansatz}
\Psi(x,E) = \Psi_{\rm in}(x,E) + \Psi_{\rm out}(x,E).
\end{equation}
For $|E|>\Delta_-$, we have $\Psi_{\rm in}\ne 0$ and normalization of the scattering state follows 
automatically by the conservation of probability current.
Subgap states with $|E|<\Delta_-$ are included in Eq.~\eqref{Ansatz} by putting $\Psi_{\rm in}=0$, where normalization is imposed by
\begin{equation}\label{normalsubgap}
\int_{-\infty}^\infty dx \, \Psi^\dagger(x,E) \cdot \Psi(x,E) =1.
\end{equation}

\emph{Continuum and mixed-character states.---}
We first consider the case $|E|>\Delta_-$, such that we have either continuum or mixed-character states and $\Psi_{\rm in}\ne 0$. With 
the length $L$ of the superconducting banks, the Heaviside step function $\Theta(x)$, and using the momenta $k_\pm$ in Eq.~\eqref{kdef},
the four possible types ($s=1,2,3,4$) of incoming bispinor states are indicated in Fig.~\ref{fig1}(a) and 
are given by
\begin{widetext}
\begin{equation} \Psi_{\rm in}(x) =  \frac{\Theta(-x)}{\sqrt{L}}
\left[ \delta_{s,1}  e^{ik_{+}x} \left( \begin{array}{l}  \psi_{+,e} \\ 0 \end{array} \right) +
\delta_{s,2} e^{ik_{-}x} \left( \begin{array}{l} 0 \\ \psi_{-,h} \end{array} \right) \right] +
\frac{\Theta(x)}{\sqrt{L}}
\left[ \delta_{s,3}e^{-ik_{+}x} \left( \begin{array}{l}  \psi_{+,h} \\ 0 \end{array} \right) +
\delta_{s,4} e^{-ik_{-}x} \left( \begin{array}{l} 0 \\  \psi_{-,e} \end{array} \right) \right].
\label{Psiin2}
\end{equation}
In particular, $s=1$ ($s=4$) refers to electron-like states with momenta near $+k_F$ ($-k_F$). 
Similarly, $s=2$ ($s=3$) describes hole-like states with momenta near $-k_F$ ($+k_F$).  Hence 
the channels $s=1,3$ describe states with index $\alpha=+$, while for $s=2,4$, we have $\alpha=-$. 
The two-spinor structure in Eq.~\eqref{Psiin2} refers to right- and left-mover space, 
with the Nambu spinors $\psi_{\alpha,e/h}(E)$ in Eq.~\eqref{tildepsi1}.
For the  outgoing state, we use the Nambu spinors $\tilde\psi_{\alpha, e/h}(E)$ in Eq.~\eqref{tildepsi1} and write
\begin{equation}
\Psi_{\rm out}(x) =  \frac{\Theta(-x)}{\sqrt{L}}
\left[ a  e^{-ik_{+}x} \left( \begin{array}{l} \tilde\psi_{+,h} \\ 0 \end{array} \right) +
b e^{-ik_{-}x} \left( \begin{array}{l} 0 \\ \tilde\psi_{-,e} \end{array} \right) \right] 
+\frac{\Theta(x)}{\sqrt{L}}
\left[ c e^{ik_{+}x} \left( \begin{array}{l}  \tilde\psi_{+,e} \\ 0 \end{array} \right) +
d e^{ik_{-}x} \left( \begin{array}{l} 0 \\  \tilde\psi_{-,h} \end{array} \right) \right]
\label{Psiout3}
\end{equation}
\end{widetext}
where the complex-valued scattering amplitudes $(a,b,c,d)$ implicitly depend on the incident-state type $s$ and its energy $E$. They are determined by solving the matching condition \eqref{BC2}.

The corresponding normalization condition follows by noting that Eqs.~\eqref{BC2} and \eqref{BdG} imply that the probability current density, 
\begin{equation}\label{contJ}
  J(x,E)=v_F\Psi_{\rm in}^\dagger \sigma_z\tau_z \Psi_{\rm in}^{} + 
    v_F\Psi_{\rm out}^\dagger \sigma_z\tau_z \Psi_{\rm out}^{},
\end{equation}
must be continuous at $x=0$.   
Using $\Psi(x,E)$ in Eqs.~\eqref{Psiin2} and \eqref{Psiout3}, the normalization 
condition for continuum and mixed-character states ($|E|>\Delta_-)$ follows with $\tilde\gamma_\alpha(E)$ in Eq.~\eqref{tildegamma} as
\begin{eqnarray}\nonumber 
&&    (|a|^2+|c|^2)\sinh\tilde\gamma_+ +(|b|^2+|d|^2)\sinh\tilde\gamma_- =\\
&&\qquad =  \frac{\delta_{s,1}+\delta_{s,3}}{\nu_+(E)} + \frac{\delta_{s,2}+\delta_{s,4}}{\nu_-(E)},
\label{normalize}
\end{eqnarray}
where we introduced a superconducting density-of-states factor for $\alpha$-movers, 
\begin{equation} \label{dosdef}
    \nu_{\alpha}(E) = \frac{|E_\alpha|}{\sqrt{E_\alpha^2-\Delta^2}}   \Theta(|E_\alpha|-\Delta) . 
\end{equation} \\
The SDE is linked to the fact that $\nu_+(E)\ne \nu_-(E)$ for $q\ne 0$.
%\footnote{For $q=0$, by writing $\tilde\gamma_\pm\to \tilde \gamma$, Eq.~\eqref{normalize} reduces to  $|a|^2+|b|^2+|c|^2+|d|^2=1/\cosh\tilde\gamma$.  The $1/\cosh\tilde\gamma$ factor 
%can be absorbed by rescaling $\tilde\psi_\alpha\to \psi_\alpha$ in $\Psi_{\rm out}$, implying the usual normalization condition $|a|^2+|b|^2+|c|^2+|d|^2=1$.}

\emph{Subgap states.---}
We now turn to subgap states with $|E|<\Delta_-$, where $\Psi_{\rm in}=0$ and the bispinor wave function $\Psi_A(x,E)$ for an Andreev state
can be obtained by analytic  continuation of Eq.~\eqref{Psiout3} with $k_\alpha(E)\to i\kappa_\alpha(E)$ as specified in Eq.~\eqref{kdef},  
\begin{widetext}
\begin{equation}
\Psi_A(x,E) =  \Theta(-x)\left[ a  \frac{e^{\kappa_{+}x}}{\sqrt{L_+}}
\left( \begin{array}{l} \tilde\psi_{+,h} \\ 0 \end{array} \right) +
b \frac{e^{\kappa_{-}x} }{\sqrt{L_-}} \left( \begin{array}{l} 0 \\ \tilde\psi_{-,e}\end{array} \right) \right] +
\Theta(x)
\left[ c \frac{e^{-\kappa_{+}x}}{\sqrt{L_+}} \left( \begin{array}{l}  \tilde\psi_{+,e}\\ 0 \end{array} \right) +
d\frac{e^{-\kappa_{-}x} }{\sqrt{L_-}} \left( \begin{array}{l} 0 \\  \tilde\psi_{-,h} \end{array} \right) \right]
\label{Psiabs}
\end{equation}
\end{widetext}
with the localization length  $L_\alpha = \frac{1}{2\kappa_\alpha}$ for $\alpha$-movers.
From Eq.~\eqref{normalsubgap}, we find that the coefficients $(a,b,c,d)$ obey the standard normalization condition $|a|^2+|b|^2+|c|^2+|d|^2=1.$

The corresponding linear algebra problem obtained from the matching conditions \eqref{BC2} can then 
be written as homogeneous $4\times 4$ matrix problem, $M (a',b',c',d')^T=0$, with rescaled amplitudes $a'=a/\sqrt{L_+}$ (and so on).
For junction transparency ${\cal T}$ (with $r=\sqrt{1-{\cal T}}$) and phase difference $\varphi$, with $\gamma_\alpha(E)=\cos^{-1}(E_\alpha/\Delta)$ in Eq.~\eqref{tildegamma}, 
the matrix $M$ is explicitly given by
\begin{equation}\label{Mdef}
    M=\left( \begin{array}{cccc} 
    -\sqrt{\cal T} & 0 & e^{\frac{i}{2}(2\gamma_+ +\varphi)} & r e^{\frac{i}{2}(\gamma_+-\gamma_-+\varphi)}\\
    -\sqrt{\cal T} & 0 & e^{-\frac{i}{2}(2\gamma_++\varphi)} & r e^{-\frac{i}{2}(\gamma_+-\gamma_-+\varphi)}\\
      0 & -\sqrt{\cal T} & r e^{\frac{i}{2}(\gamma_+-\gamma_-+\varphi)} & e^{-\frac{i}{2} (2\gamma_--\varphi)} \\ 
     0 & -\sqrt{\cal T} & r e^{-\frac{i}{2}(\gamma_+-\gamma_-+\varphi)} & e^{\frac{i}{2} (2\gamma_--\varphi)} 
    \end{array}\right).
\end{equation}
The spectral condition for Andreev states  follows from the vanishing of the determinant, 
\begin{equation}\label{ABS0}
    \sin^2\left( \frac{\gamma_++\gamma_-}{2}\right) = {\cal T} \sin^2\left( \frac{\gamma_+-\gamma_- + \varphi}{2}\right).
\end{equation}
The corresponding eigenvectors of $M$ determine the coefficients $(a,b,c,d)$ and thus the associated normalized
spinor eigenstate  \eqref{Psiabs}.

We note that for ${\cal T}<1$, Eq.~\eqref{ABS0} differs from 
the corresponding result of Ref.~\cite{Davydova2022}, 
which also does not recover the $q=0$ Andreev state dispersion  \cite{Furusaki1991,Beenakker1991}, 
\begin{equation}\label{q=0}
E^{(0)}_\pm(\varphi)=\pm \Delta \sqrt{1-{\cal T}\sin^2(\varphi/2)} .
\end{equation}  
The discrepancy is due to a technical mistake in Ref.~\cite{Davydova2022} which has been corrected in Ref.~\cite{FuErratum}.
For small CP momentum, $q\xi \ll 1$, by expanding Eq.~\eqref{ABS0} up to terms of order ${\cal O}( (q\xi)^2 )$, we find the  
Andreev state dispersion 
\begin{equation}\label{smallq}
    E_\pm(\varphi) =E_\pm^{(0)}(\varphi) - q\xi \sqrt{\cal T} 
\frac{\cos(\varphi/2){\rm sgn }(\sin(\varphi/2))}
    { E_\pm^{(0)}(\varphi) } .
\end{equation} 
In the ballistic limit of perfect transparency ${\cal T}=1$,  Eq.~\eqref{smallq} turns out to be exact for 
all values of $q\xi$, see  Sec.~\ref{sec3b} below, in accordance with the results of Ref.~\cite{Davydova2022}.

%%%%%%%%%%%%%%%%%%%%%%%%%%%%%%%%%%%%%%%%%%%%%%%%%%%%%%%%%%%%%%%%%%%%%%%%%%%%%%%%%%%%%%%%%%
\section{Equilibrium superconducting diode effect}\label{sec3}

In this section, we discuss the equilibrium CPR for the superconducting weak link defined by the BdG problem \eqref{BdG}, in particular how the SDE efficiency $\eta_0$  
  depends on the CP momentum $2q$ and on the junction transparency ${\cal T}$.
We start by deriving general expressions for the CPR in Sec.~\ref{sec3a}.  We then continue with
 the ballistic case ${\cal T}=1$ in Sec.~\ref{sec3b}, where we reproduce the 
results of Ref.~\cite{Davydova2022}.  The case ${\cal T}<1$ is  addressed in Sec.~\ref{sec3c}.  
In addition, in App.~\ref{appA}, we discuss a different Josephson diode model,
where the weak link is represented by a spin-orbit coupled nanowire in a magnetic Zeeman field.
In that case, a magnetochiral anisotropy is responsible for the SDE.

\subsection{Current-phase relation}\label{sec3a}

The equilibrium supercurrent $I(\varphi)= 
I_{\rm cont}+I_{A}$ can conveniently be evaluated at $x=0$,  
\begin{equation}\label{totalCPR}
    I= ev_F \sum_E n_F(E) \left. \Psi^\dagger(x,E) \sigma_z \Psi(x,E)\right|_{x=0},   
\end{equation}
with the Fermi function $n_F(E)=1/(e^{E/k_BT}+1)$ at temperature $T$ (with the Boltzmann 
constant $k_B$). 
The absence of $\tau_z$ in Eq.~\eqref{totalCPR} as compared to 
Eq.~\eqref{contJ} for the probability current $J(x,E)$ arises because we study the charge current. 
Since $U^\dagger \sigma_z U=\sigma_z$, the  
transformation \eqref{Ugauge} has no effect in Eq.~\eqref{totalCPR}.
Current contributions from continuum and mixed-character states with $|E|>\Delta_-$ are summarized in 
$I_{\rm cont}$, and those from Andreev subgap states in $I_A$.  We separately address both terms.

\emph{Continuum contribution.---} Using Eqs.~\eqref{Psiin2} and \eqref{Psiout3},  taking the limit $L\to \infty$
and using the density-of-states factors $\nu_\alpha(E)$ in Eq.~\eqref{dosdef}, we obtain 
\begin{equation} \label{Icontn}
   I_{\rm cont} = e \sum_{s=1}^4  \int_{-\infty}^\infty \frac{dE}{2\pi}\, n_F(E) \nu_{\alpha_s}(E) I_s(E),
\end{equation}
with $\alpha_{s}=(-1)^{1+s}$ for incident-state type $s$, see Eq.~\eqref{Psiin2}, and 
\begin{equation} \label{Isdef}
I_s(E) =\cosh (\tilde\gamma_{\alpha_s})  \times 
\left\{ \begin{array}{cc} 
 |c|^2-|d|^2  ,& s=1,2, \\  |a|^2-|b|^2,& s=3,4.\end{array}\right.
\end{equation}
with $\tilde\gamma_\alpha(E)$ in Eq.~\eqref{tildegamma}.
The scattering amplitudes $(a,b,c,d)$, see Eq.~\eqref{Psiout3}, and hence also $I_s(E)$, follow by solving the linear 
algebra problem posed by the matching conditions \eqref{BC2}.  

\emph{Subgap contribution.---} The Andreev current follows from Eq.~\eqref{totalCPR} as
\begin{eqnarray}\nonumber
    I_A &=&e v_F\sum_\lambda n_F(E_\lambda) \left( \frac{|a_\lambda|^2}{L_+}- \frac{|b_\lambda|^2}{L_-} \right)\\
    &=&ev_F \sum_\lambda n_F(E_\lambda) \left( \frac{|c_\lambda|^2}{L_+}- \frac{|d_\lambda|^2}{L_-} \right), \label{Iabs0}
\end{eqnarray}
where $E_\lambda(\varphi)$ is the dispersion relation of the respective Andreev state and the
amplitudes $(a,b,c,d)$ in Eq.~\eqref{Psiabs} follow from the corresponding eigenvector of the matrix $M$ in Eq.~\eqref{Mdef},
with $L_\alpha=\frac{1}{2\kappa_\alpha}$.

\subsection{Perfect transparency}\label{sec3b}

We first briefly demonstrate that in the ballistic limit with ${\cal T}=1$,  our formalism reproduces the results
of Ref.~\cite{Davydova2022}.  For ${\cal T}=1$, the matching condition \eqref{BC2}
simplifies to $\Psi(0^-)=e^{i\tau_z \varphi/2} \Psi(0^+)$, and thus chirality is conserved, with eigenvalue $\alpha=\pm$ of $\sigma_z$.
 Mixed-character states with energies in the window $\Delta_-\le |E|\le \Delta_+$ then split up into 
Andreev states and continuum states without evanescent contribution. Both types of states 
coexist in this spectral range for ${\cal T}=1$. 

Let us begin with subgap states with $|E_\alpha|<\Delta$,  where insertion of Eq.~\eqref{Psiabs} into
the matching conditions gives decoupled linear equations for the scattering amplitudes.  Using
the notation $\hat \gamma_\alpha=\gamma_\alpha+\alpha\frac{\varphi}{2}$, we find from Eq.~\eqref{Mdef} the relations
\begin{equation}\label{MMM}
    a= c e^{i\hat \gamma_+}= c e^{-i\hat \gamma_+},\qquad
    b= d e^{i\hat \gamma_-}= d e^{-i\hat \gamma_-}.
\end{equation}
The resulting dispersion equation, $e^{2i\hat \gamma_\alpha}=1$, agrees with Eq.~\eqref{ABS0}.
The solution is given by $\hat\gamma_\alpha=\pi n$ with integer $n$.     Since $\gamma_\alpha(E) \in (0,\pi)$, there
are exactly two solutions corresponding to $\alpha=\pm$. 
Defining $E_{A}(\varphi) = \Delta \cos(\varphi/2) -  v_Fq$, we obtain for $\alpha=+$ the 
Andreev level energy $E_1=-E_A$ with $a=-c=\frac{1}{\sqrt2}$ and $b=d=0$, see Eq.~\eqref{MMM}. Similarly, for $\alpha=-$, we find the energy $E_2=+E_A$ 
with $b=d=\frac{1}{\sqrt2}$ and $a=c=0$. 
These energies agree with the small-$q$ Andreev dispersion in Eq.~\eqref{smallq} for ${\cal T}=1$.
We thus conclude that Eq.~\eqref{smallq} is exact for all values of $q\xi$ in the ballistic limit.
The Andreev current follows from Eq.~\eqref{Iabs0} as  
\begin{equation} \label{Iabs}
    I_{A}  = e\Delta \sin(\varphi/2)\, \tanh \left( \frac{ \Delta \cos(\varphi/2) -  v_Fq}{2k_B T}\right).
\end{equation}

Next we turn to continuum states with $|E_\alpha|>\Delta$.  For the four possible incident-state types,
the matching conditions give the nonvanishing scattering probabilities  
\begin{eqnarray}\nonumber
    |c_{s=1}|^2&=& 1-|a_{s=1}|^2 = \frac{\Delta^2}{E_{\alpha=+}^2 - \Delta^2 \cos^2(\varphi/2)} , \\ 
   |d_2|^2 &=& 1-|b_2|^2= \frac{\Delta^2}{E_{\alpha=-}^2 - \Delta^2 \cos^2(\varphi/2)} , \\
  |d_3|^2&=& 1-|c_3|^2 = |c_1|^2 ,\quad
     |b_4|^2= 1-|d_4|^2 = |d_2|^2.\nonumber
\end{eqnarray} 
From Eq.~\eqref{Icontn}, we  obtain  
\begin{equation}\label{Icont2}
    I_{\rm cont}=e\sum_{\alpha=\pm }\int \frac{dE}{\pi} n_F(E) \frac{\alpha |E_\alpha|\sqrt{E_\alpha^2-\Delta^2}}{E_\alpha^2-\Delta^2\cos^2(\varphi/2)} \Theta(|E_\alpha|-\Delta).
\end{equation}
In general, this result depends on the phase difference $\varphi$.  
For $q=0$, we have $E_\alpha=E$ and Eq.~\eqref{Icont2} implies $I_{\rm cont}=0$.

\emph{Zero-temperature limit.---} Putting $T=0$, the continuum contribution  \eqref{Icont2} is given by
\begin{eqnarray} \nonumber
I_{\rm cont} &=& e \lim_{\Omega\to \infty} \left( \int_\Delta^{\Omega+v_Fq}-\int_\Delta^{\Omega-v_Fq} \right) \frac{dE}{\pi} \\  &\times&
 \frac{E\sqrt{E^2-\Delta^2}}{E^2-\Delta^2\cos^2 (\varphi/2)} = e\Delta \frac{2q\xi}{\pi}. \label{Icontball}
 \end{eqnarray}
Note that the integration can be carried out in an almost trivial manner by writing it in the above form, 
where we observe that the phase dependence drops out in the $T=0$ limit.
One can now easily read off the critical currents $I_{c\pm}$ which determine the SDE efficiency according to 
Eq.~\eqref{eta0},
\begin{equation}\label{eta0T1}
  \eta_0 = 1 - \frac{2-4q\xi/\pi}{ 1+\sqrt{1-(q\xi)^2}}.
\end{equation}
For $q\xi\ll 1$, Eq.~\eqref{eta0T1} gives $\eta_0\approx \frac{2}{\pi} q\xi$, while for $q\xi\to 1$, we obtain
$\eta_0\approx \frac{4}{\pi}-1\approx 0.27$.  The point of maximal efficiency is reached for
$q\xi\approx 0.9$ with $\eta_0\approx 0.4$. 

\begin{figure}
\centering
\includegraphics[width=0.49\textwidth]{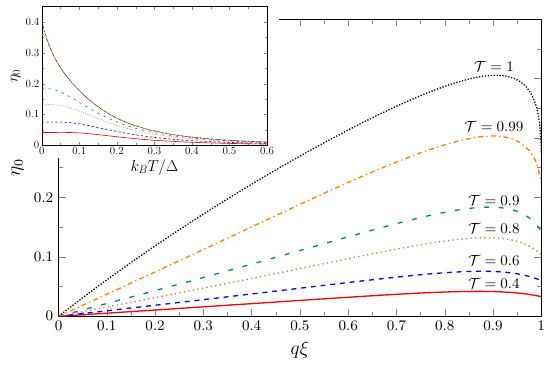}
\caption{SDE efficiency $\eta_0$ in Eq.~\eqref{eta0} corresponding to the equilibrium CPR, computed from Eqs.~\eqref{Icontn} and \eqref{Iabs0}.
The main panel shows $\eta_0$ vs the CP momentum parameter $q\xi$ for several junction transparencies ${\cal T}$ in the zero-temperature ($T=0$) limit.
The ${\cal T}=1$ result is given by Eq.~\eqref{eta0T1}.
The inset shows $\eta_0$ vs temperature $T$ (in units of $\Delta/k_B$) for the same transparencies as in the main panel, choosing  $q\xi=0.9$ where near-optimal 
    SDE efficiency is realized for $T=0$.  }    \label{fig2}
\end{figure}

\subsection{Non-ideal transparency} \label{sec3c}

We now turn to numerical results for ${\cal T}<1$, where our code accurately reproduces the 
exact results for ${\cal T}=1$ in Sec.~\ref{sec3b}.   Another check passed by our code is given by the small-$q$ 
expansion of the Andreev level energies in Eq.~\eqref{smallq}.   
In numerical calculations, we obtain the CPR from Eqs.~\eqref{Icontn} and \eqref{Iabs0} after solving the 
corresponding matching problem for a given parameter set $(q\xi,{\cal T},k_BT/\Delta)$.  
Given the CPR, we then read off the critical currents $I_{c\pm}$ and determine the SDE efficiency $\eta_0$ according to Eq.~\eqref{eta0}. 

The main panel in Fig.~\ref{fig2} is for the zero-temperature limit and shows the dependence of $\eta_0$ on the CP momentum parameter
$q\xi$ for selected transparencies ${\cal T}$, including the ballistic case where Eq.~\eqref{eta0T1} applies.  Our results for ${\cal T}<1$ coincide with the 
recently corrected \cite{FuErratum} version of Ref.~\cite{Davydova2022}.
We conclude that for decreasing transparency ${\cal T}$, the SDE efficiency also decreases since the anharmonic content in the CPR becomes suppressed.
For a given ${\cal T}$, we observe that the optimal efficiency is reached for $q\xi\approx 0.9$, practically independent of the value of ${\cal T}$.

The inset of Fig.~\ref{fig2} shows the impact of thermal fluctuations on the SDE efficiency for the near-optimal case $q\xi=0.9$ and 
the same transparencies ${\cal T}$ as in the main panel.   We observe a rapid decrease of $\eta_0$ with $k_BT/\Delta$ for transparencies near the ballistic limit, 
while for ${\cal T}\alt 0.9$, the efficiency approximately remains at the $T=0$ value for $k_BT \alt 0.1\Delta$.  However, 
with further increase of the temperature, the efficiency $\eta_0$ becomes very small for all values of ${\cal T}$.

%%%%%%%%%%%%%%%%%%%%%%%%%%%%%%%%%%%%%%%%%%%%%%%%%%%%%%%%%%%%%%%%%%%%%%%%%%%%%%%%%%%%%%%%%%%%%
\section{NS junction}\label{sec4}

In this section, we discuss charge transport through a voltage-biased NS junction \cite{Blonder1982} of transparency ${\cal T}$ between
 a normal metal ($x<0$) and a helical superconductor $(x>0)$ with finite CP momentum $2q$.  
We derive general expressions for the nonlinear differential conductance $G(V)$ in Sec.~\ref{sec4a}.
Analytical results for the conductance in the ballistic limit  $({\cal T}=1)$ are summarized in Sec.~\ref{sec4b}.  
For non-ideal transparency ${\cal T}<1$, we present numerical results for $G(V)$ in Sec.~\ref{sec4c}.

\subsection{Nonlinear conductance}\label{sec4a}

We describe the normal-conducting region ($x<0$) by the Hamiltonian $H_L$ in Eq.~\eqref{HLR} but with vanishing pairing gap ($\Delta = 0$). 
Including the bias voltage $V$ as potential shift in the normal region $x<0$, 
it is convenient to perform a gauge transformation, $\Psi(x,t) \mapsto e^{i\tau_z Vt} \Psi(x,t)$ for $x<0$. 
For an incoming state with energy $E$, one thereby arrives at an effectively stationary scattering problem
characterized by the matching condition
\begin{equation}\label{BCE}
\Psi(0^-,E) = \frac{1}{\sqrt{\cal T}} (\sigma_0+r\sigma_x)  \tau_0  \Psi(0^+,E).
\end{equation}
For a calculation of the $I$-$V$ curve, the bias voltage is then accounted for
by $V$-dependent Fermi factors for incoming electrons or holes in the normal lead.
Using the scattering states in Eqs.~\eqref{Psiin2} and \eqref{Psiout3}, the bispinors at $x=0^\pm$ are given by
\begin{eqnarray}
    \Psi(0^-,E) &=& \left(\begin{array}{c} \delta_{s,1} \chi_e \\ \delta_{s,2} \chi_h\end{array}\right) 
    + \left(\begin{array}{c} a\chi_h \\ b \chi_e\end{array}\right) ,\\
    \nonumber
    \Psi(0^+,E) &=& \left(\begin{array}{c} \delta_{s,3} \psi_{+,h}(E) \\ \delta_{s,4} \psi_{-,e}(E) \end{array}\right) +
    \left(\begin{array}{c}  c\tilde\psi_{+,e}(E) \\ d \tilde\psi_{-,h}(E) \end{array}\right) ,
\end{eqnarray}
with Nambu spinors $\psi_{\alpha,e/h}$ for incoming states  
and $\tilde\psi_{\alpha,e/h}$ for outgoing states, see Eq.~\eqref{tildepsi1}.  
The corresponding spinors in the normal lead follow accordingly by
letting $\Delta\to 0$, 
\begin{equation}
    \chi_e=\left(\begin{array}{c} 1\\ 0\end{array}\right) ,
\quad \chi_h= \left(\begin{array}{c} 0 \\1\end{array}\right) .
\end{equation}
The scattering amplitudes $(a,b,c,d)$ depend on the incoming state type $s$ and are determined by the matching condition \eqref{BCE}. 

Assuming that the density of states in the normal region is constant and given by the normal-phase value on the superconducting side, $\nu_N=1/(2\pi v_F)$ per spin,  the DC current-voltage characteristics, $I(V)$, 
then follows from Eq.~\eqref{Icontn}.   With $\tilde\gamma_\alpha(E)$ in Eq.~\eqref{tildegamma} and $\nu_\alpha(E)$ in Eq.~\eqref{dosdef}, we obtain
\begin{eqnarray} \label{IVNS}
   I(V) &=& e\sum_{s=1}^4  \int_{-\infty}^\infty \frac{dE}{2\pi} \tilde N_s(E) \tilde I_s(E),
 \\ \nonumber
     \tilde N_s(E) &=& \left\{\begin{array}{cc} n_F(E-eV),&s=1,\\
     n_F(E+eV),&s=2,\\ n_F(E)\nu_+(E),& s=3,\\ n_F(E)\nu_-(E),& s=4,\end{array}\right. \\
     \tilde I_s(E) &=& \left\{ \begin{array}{cc}
 \cosh(\tilde\gamma_+)|c|^2-\cosh(\tilde\gamma_-) |d|^2 ,& s=1,2,\\   \nonumber
 |a|^2-|b|^2,& s=3,4.\end{array} \right. 
\end{eqnarray} 
The differential conductance is thus given by
\begin{eqnarray}\nonumber
   G(V)  &=& \frac{dI}{dV}= e \int \frac{dE}{2\pi} \Biggl( \frac{dn_F(E-eV)}{dV} \tilde I_{1}(E) +\\
 &&\qquad +\frac{dn_F(E+eV)}{dV} \tilde I_{2}(E)  \Biggr),\label{condNS}
\end{eqnarray}
where only the scattering channels $s=1,2$ contribute in the chosen gauge. 

In what follows, we study the nonlinear NS conductance in the zero-temperature limit ($T=0$).  
Using the conductance quantum $G_0=2e^2/h$, Eq.~\eqref{condNS} then simplifies to
\begin{equation}\label{zeroT}
\frac{G(V)}{G_0}= \frac12 \left( \tilde I_{1}(eV)-\tilde I_{2}(eV) \right).
\end{equation}  
Importantly, one finds from Eq.~\eqref{IVNS} that the NS conductance \eqref{zeroT} is always even under voltage reversal, $G(-V)=G(V)$.  NS junctions involving a helical superconductor therefore do not exhibit rectification.  

\begin{figure*}
\centering
  \includegraphics[width=\textwidth]{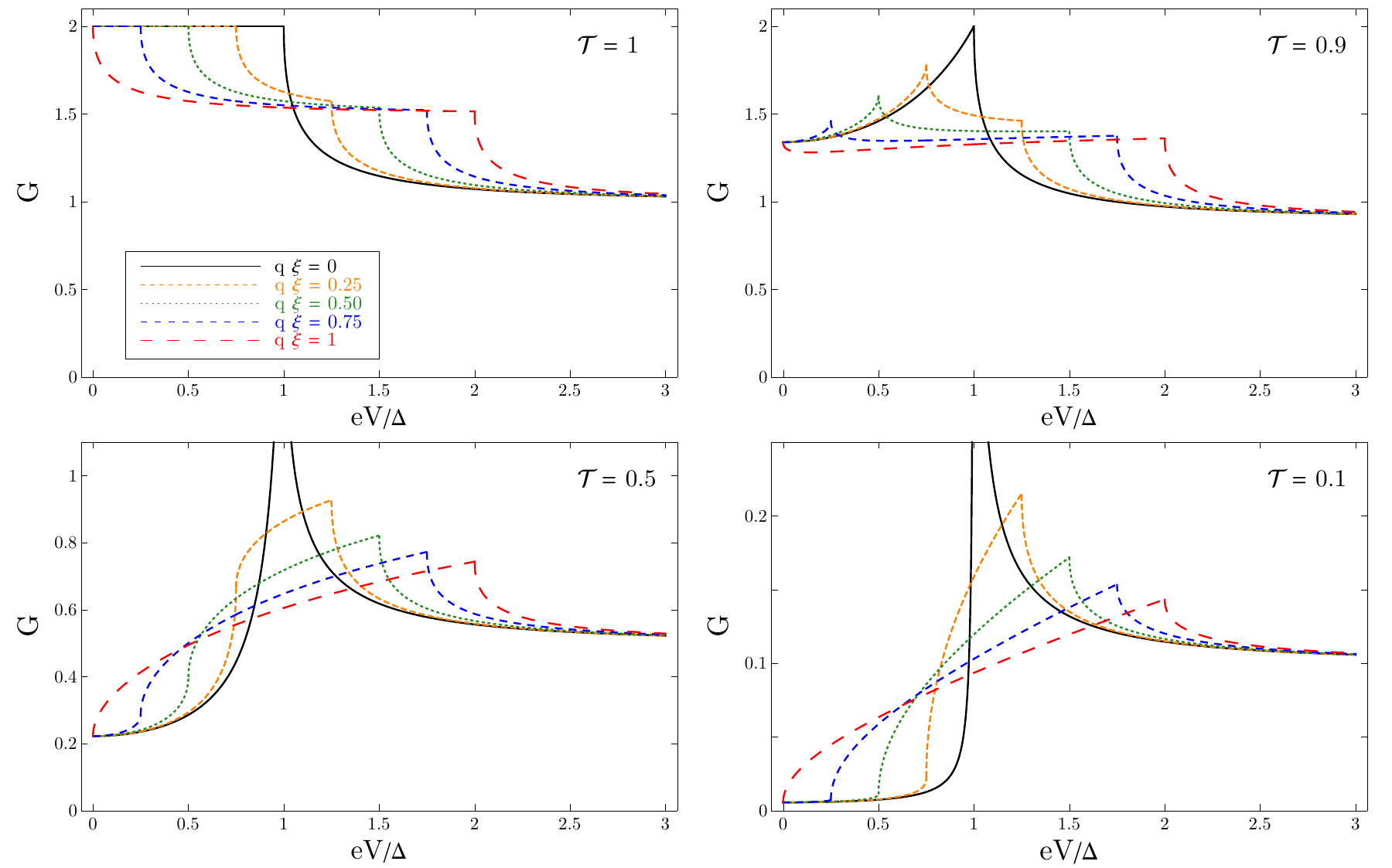}
\caption{Zero-temperature nonlinear conductance $G(V)$ (in units of $G_0=2e^2/h$)  
for a NS junction with several values for the CP momentum parameter $q\xi$.
Since $G(-V)=G(V)$, we only show $G(V)$ for positive voltages.
Panel (a) is for transparency ${\cal T}=1$ and was obtained from Eq.~\eqref{GT1}.
Panels (b), (c), and (d) are for ${\cal T}=0.9, 0.5, 0.1$, respectively, 
and were obtained by numerically solving Eqs.~\eqref{IVNS} and \eqref{zeroT}. 
}
\label{fig3}
\end{figure*}

\subsection{Ballistic limit}\label{sec4b}

We first consider the case  ${\cal T}=1$, where the matching condition \eqref{BCE} is 
fulfilled automatically by continuous bispinor wave functions $\Psi(x,E)$.
The non-vanishing scattering amplitudes ($a_s,b_s,c_s,d_s$) for given energy $E$ and scattering channel $s$ 
are given by
\begin{eqnarray}
a_1&=& \sigma_+ e^{-\tilde\gamma_+},\quad c_1= \sqrt2 ,\\ 
\nonumber b_2&=& \sigma_- e^{-\tilde\gamma_-},\quad d_2=\sqrt2\sigma_- ,\\
\nonumber a_3 &=& \sigma_+ \frac{1-e^{-\tilde\gamma_+}}{ \sqrt{2\cosh\tilde\gamma_+}},\quad
c_3=-\frac{e^{-\tilde\gamma_+}}{\sqrt{\cosh\tilde\gamma_+}},\\
\nonumber b_4&=& \frac{1-e^{-\tilde\gamma_-}}{\sqrt{2\cosh\tilde\gamma_-}} ,\quad
d_4 = -\frac{e^{-\tilde\gamma_-}}{\sqrt{\cosh\tilde\gamma_-}} ,
\end{eqnarray}
with $\sigma_\alpha={\rm sgn}(E_\alpha)$.
The zero-temperature nonlinear conductance \eqref{zeroT} thus follows from Eq.~\eqref{IVNS} as
\begin{eqnarray}\nonumber
    \frac{G(V)}{G_0}&=& 1+\frac12\sum_{\alpha=\pm} e^{-2\tilde \gamma_\alpha(eV)} \\
       &=&1+\frac12 \left(|a_1(eV)|^2+|b_2(eV)|^2\right) ,
\end{eqnarray}
where $|a_1|^2$ and $|b_2|^2$ are the probabilities for electron- and hole-type Andreev reflections, respectively, where the reflected quasiparticle stays
on the same branch ($\alpha=\pm$) as the incident one.
Using Eq.~\eqref{tildegamma}, we obtain $G$ as a function of the dimensionless parameters $v=eV/\Delta$ and
$q\xi$ in explicit form,
\begin{eqnarray}  \label{GT1}
  \frac{G(v,q\xi)}{G_0} &=& 1+ \frac12 \sum_{\alpha=\pm}\Biggl(
  \Theta(1-|v-\alpha q\xi|) + \\ 
  \nonumber &+&   
\frac{\Theta(|v-\alpha q\xi|-1)}{\left(|v-\alpha q\xi|+ 
\sqrt{\left(v-\alpha q\xi\right)^2-1}\right)^2} \Biggr).
\end{eqnarray}
Note that $G_0\le G\le 2G_0$.  

In Fig.~\ref{fig3}(a), we show the nonlinear conductance \eqref{GT1}
for  different values of the CP momentum parameter $q\xi$,
restricting ourselves to positive voltages in view of the symmetry relation $G(-V)=G(V)$.
For ${\cal T}=1$, $G(V)$ stays at the maximal value $G=2G_0$ for 
voltages below the lower spectral gap, $eV<\Delta_-=\Delta(1-q\xi)$.  
The nonlinear conductance then monotonically decreases with increasing $V$ for $eV>\Delta_-$,
and approaches the limiting value $G=G_0$ for very large voltage $eV\gg \Delta_+=\Delta(1+q\xi)$. 
Clearly, the derivative $\frac{dG}{dV}=\frac{d^2I}{dV^2}$ shows discontinuous jumps 
at the voltages $eV=\Delta_\pm$ corresponding to the two spectral gaps, where the
conductance decreases in a step-like way.

\subsection{Non-ideal transparency}\label{sec4c}

We next turn to the case of non-ideal junction transparency, ${\cal T}<1$,
where we numerically solve the matching problem described in Sec.~\ref{sec4a} 
and determine the $T=0$ conductance \eqref{zeroT} from Eq.~\eqref{IVNS}.
The dependence of the conductance   on  
$q\xi$ is shown for several   ${\cal T}$ in Fig.~\ref{fig3}.
We draw several conclusions:
\begin{itemize}
\item For large voltage $V\to \infty$, the NS conductance  $G(V)$
approaches the $q$-independent Ohmic conductance \cite{Nazarov2009}
\begin{equation}\label{ohm}
    G_\infty={\cal T} G_0
\end{equation}
of the corresponding NN contact. 
\item The conductance also becomes independent of the CP momentum $2q$ 
in the linear-response limit $V\to 0$, where the conductance scales as 
$G(0)\propto {\cal T}^2$ because it is governed by Andreev reflections. 
The conductance $G(V)$ thus depends on $q\xi$ only for voltages that are 
neither very small nor very large.
\item  For the conventional case $q=0$, we recover a sharp universal
conductance peak with $G=2G_0$ for $eV=\Delta$ \cite{Blonder1982}, 
with a discontinuity in the  derivative $dG/dV$. 
\item For finite CP momentum $2q\ne 0$, 
the voltage-dependent conductance has \emph{two} singular features:  
the single conductance peak at $eV=\Delta$ found for $q=0$ splits into
a pair of singularities at the Doppler shifted spectral gaps, $eV=\Delta_\pm$, as
in the ballistic case. At these voltages, the derivative $dG/dV$ is discontinuous
since a peak in the $\alpha$-dependent superconducting density of states 
\eqref{dosdef} aligns with an effective Fermi level on the normal side.  
The doubling of singular features for $q\ne 0$ is a characteristic prediction for 
the NS conductance of a helical superconductor. 
\item For $q\ne 0$, the peak conductance is found at $eV=\Delta_-$ 
if the transparency is large, with a step-like conductance decrease at $eV=\Delta_+$,
see Fig.~\ref{fig3}(b).
On the other hand, for low transparency, there is a step-like conductance increase at $eV=\Delta_-$ and the peak instead occurs at $eV=\Delta_+$, see Figs.~\ref{fig3}(c,d).  
\item For $q\ne 0$, the peak conductance value is not universal anymore and
 explicitly depends both on ${\cal T}$ and on $q\xi$. In particular, it
 always stays below the maximal value $2G_0$.
\end{itemize}
The observation of the above predictions in tunneling spectroscopy experiments
could help in identifying helical superconductors.

%%%%%%%%%%%%%%%%%%%%%%%%%%%%%%%%%%%%%%%%%%%%%%%%%%%%%%%%%%%%%%%%%%%%%%%%%%%%%%%%%%%%%%%%%
\section{Voltage-biased Josephson diode}\label{sec5}

We now return to the case of a Josephson diode
and study the $I$-$V$ characteristics under a constant bias voltage $V$. 
To that end, we formulate a scattering theory accounting for MAR processes.  
In contrast to previous work, see, e.g., Refs.~\cite{Bratus1995,Averin1995,Zazunov2006},
our theory allows for a finite CP momentum $2q$.

Assuming a symmetric voltage bias, $V_L=-V_R=V/2$, the phase difference is given by $\varphi(t)=2eVt$ and Eq.~\eqref{BC2} leads to the 
time-dependent matching condition
\begin{equation}\label{BCtime}
\Psi_E(0^-,t) = \frac{1}{\sqrt{\cal T}} (\sigma_0+r\sigma_x)   e^{i\tau_z eV t}\Psi_E(0^+,t),
\end{equation}
where $E$ is the energy of the incident state. We require $|E_\alpha|>\Delta$ for an incoming $\alpha$-mover.
However, the energy of outgoing states may differ from $E$
due to the absorption or emission of $eV$ quanta. In general, the energy could be of the form 
\begin{equation}\label{En}
    E_n = E+neV, \qquad ({\rm integer}\, n),
\end{equation} 
which may include subgap energies.

In Sec.~\ref{sec5a}, we describe the MAR theory for helical superconductors
($q\ne 0$), using the model in Sec.~\ref{sec2}.  Our theory recovers the results of Ref.~\cite{Averin1995}
for $q=0$. We present an analytical solution for the 
$I$-$V$ curve in the ballistic limit ${\cal T}=1$ in Sec.~\ref{sec5b}, where we
also discuss the higher harmonic components of the AC current at low voltages.
For ${\cal T}<1$, we discuss numerical results for $I(V)$ in Sec.~\ref{sec5c} and interpret 
them in terms of the MAR ladder picture  in Fig.~\ref{fig4}.
In Sec.~\ref{sec6}, we then describe the corresponding
results for the voltage-dependent efficiency $\eta(V)$ in Eq.~\eqref{efficiency}.    

\subsection{MAR theory}\label{sec5a}

\begin{figure}
\includegraphics[width=0.49\textwidth]{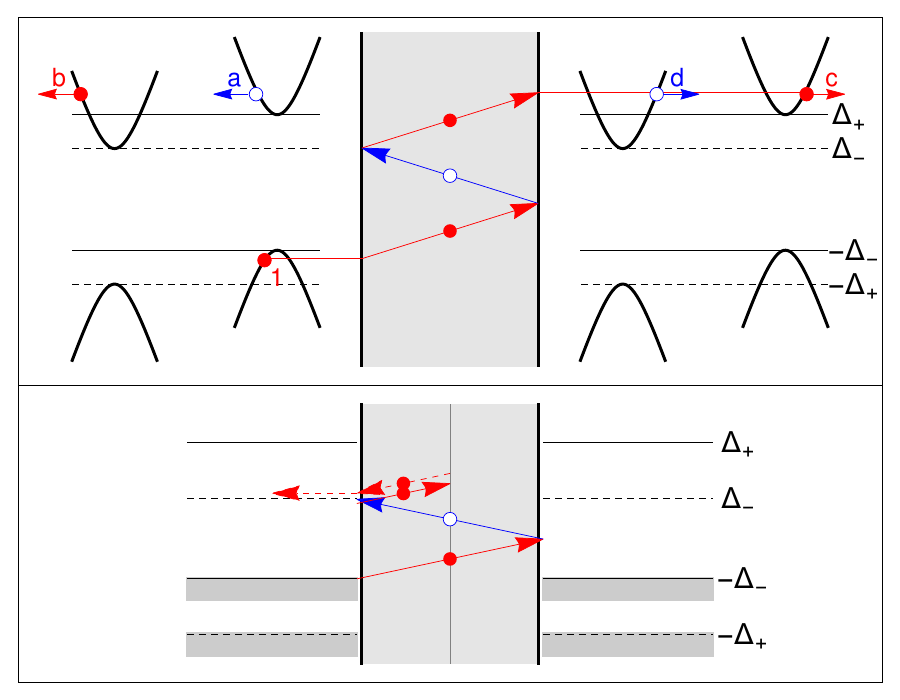} 
\caption{Sketch of the BdG dispersion and the MAR ladder picture.
Upper panel:  The quasiparticle dispersion is shown on the 
left and on the right side, where the shaded central region 
refers to the normal weak link. 
Electron $(e)$ and hole $(h)$ states are indicated by filled red (open blue) 
circles and their trajectories in energy space are shown as red (blue) lines.  
In this representation, the voltage drops across the normal region and 
the shown $e/h$ trajectories gain energy $eV$ by traversing it.
A typical MAR trajectory for ${\cal T}=1$ is shown, see also Fig.~\ref{fig1}, where an
$e$ state impinges from the left side (channel type $s=1$) with energy $E\alt -\Delta_-$. The scattering amplitudes $(a,b,c,d)$ in Eq.~\eqref{MARAnsatz} are also indicated. 
Lower panel:
In the presence of normal reflection $r\ne 0$, i.e., for ${\cal T}<1$, new types of trajectories
are possible which generate additional MAR features for $2\Delta_\pm/eV=n$ (integer $n$).
In the shown example, the voltage is $eV=\Delta_-$. In the last step, the 
electron undergoes normal reflection and ends in the left side at 
energy $E\approx \Delta_-$.  This trajectory is of first order in $r$ and causes a MAR feature at $2\Delta_-/eV=2$.
}
\label{fig4}
\end{figure}

We now construct scattering states for the finite-voltage case taking into account MAR processes.  Typical MAR trajectories in energy space (``MAR ladder'') are shown in Figs.~\ref{fig1} and \ref{fig4}.  We consider an incident $\alpha$-mover which is an electron or hole like quasiparticle with energy $E$ in the respective continuum, $|E_\alpha|>\Delta$. 
For each step of the MAR ladder, the energy of an electron changes by 
$\pm eV$ for right- or left-movers 
when traversing the normal junction region, and similarly the energy shift for holes is $\mp eV$.
The energy $E_n=E+neV$ of an outgoing (reflected or transmitted) state may therefore involve the 
emission or absorption of an integer number $n$ of $eV$ quanta.  

In what follows, we employ the electron-type Nambu spinor $\tilde\psi_{\alpha,n}(E)$ for an outgoing $\alpha$-mover with energy $E_n$. 
Similarly, we define the corresponding incoming Nambu spinors $\psi_{\alpha,n}(E)$, which are
only needed for $n=0$.
Using Eqs.~\eqref{psidef} and \eqref{tildepsi}, 
\begin{equation}\label{outgoingMAR}
    \psi_{\alpha,n}(E)=\psi_e(E_n-\alpha v_Fq),\quad 
    \tilde \psi_{\alpha,n}(E)=\tilde\psi_e(E_n-\alpha v_F q).
\end{equation}
The respective hole-type Nambu spinors follow by acting with $\tau_x$ on the spinors \eqref{outgoingMAR}.  For instance, in the new notation, the Nambu spinors $\psi_{\alpha,e/h}(E)$ in Sec.~\ref{sec2} are given by $\psi_{\alpha,0}(E)$ and $\tau_x \psi_{\alpha,0}(E)$, respectively.
Using the general scattering states in Eqs.~\eqref{Psiin2} and \eqref{Psiout3}
with four possible types $s\in\{1,2,3,4\}$ of incident states, 
the bispinor states at $x=0^\pm$ for the voltage-biased case then have the form
\begin{eqnarray} \nonumber
\Psi_E(0^-,t) &=&e^{-iEt}  \left( \begin{array}{c} \delta_{s,1}\, \psi_{+,0}(E) \\ \delta_{s,2} \,
\tau_x \psi_{-,0}(E)\end{array} \right) \\
&+&\nonumber \sum_n e^{-iE_n t}\left( \begin{array}{c} a_n\tau_x \tilde\psi_{+,n}(E) \\ b_n \tilde\psi_{-,n}(E)\end{array} \right), \\ \label{MARAnsatz}
\Psi_E(0^+,t) &=&e^{-iEt}  \left( \begin{array}{c} \delta_{s,3} \,\tau_x\psi_{+,0}(E) \\ \delta_{s,4} \,\psi_{-,0}(E)\end{array} \right)\\
&+ &\nonumber\sum_n e^{-iE_n t}
\left( \begin{array}{c} c_n \tilde\psi_{+,n}(E) 
\\ d_n \tau_x\tilde\psi_{-,n}(E)\end{array} \right).
\end{eqnarray}
We keep the dependence of the scattering amplitudes $(a_n,b_n,c_n,d_n)$ 
on the incident-state energy $E$ and channel type $s$ implicit and
define the quantities
\begin{equation}\label{rhoan}
    \rho_{\alpha,n}= \rho(E_n-\alpha v_Fq), 
\end{equation}
with $\rho(E)$ in Eq.~\eqref{tildegamma} and
\begin{equation}\label{lamdaan}
    \lambda_\alpha=\lambda(E-\alpha v_Fq) = \sqrt{\frac{2}{1+\rho^2_{\alpha,0}}}.
\end{equation}
We refer to Fig.~\ref{fig4} for an illustration of these scattering states.

To proceed, we write the matching condition \eqref{BCtime} for given incident-state energy $E$ 
separately for the upper (electron-like) and lower (hole-like) components of the Nambu spinors.  
Consider first the upper components for $s=1,2$, where we obtain a recurrence relation for spinors in 
right-left mover space,
\begin{eqnarray}
 &&   \delta_{n,0}  \left(\begin{array}{c} \delta_{s,1}\,\lambda_+\\
    \delta_{s,2} \, \rho_{-,0} \lambda_- \end{array}\right)+ 
    \left(\begin{array}{c} \rho_{+,n} a_n\\b_n\end{array}\right)  \nonumber\\ \label{s12up}&&=
  \frac{\sigma_0+r\sigma_x}{\sqrt{\cal T}}  \left(\begin{array}{c} c_{n+1}\\\rho_{-,n+1}d_{n+1}\end{array}\right).
\end{eqnarray}
Similarly, the lower components for $s=1,2$ imply the relation
\begin{eqnarray}\nonumber
 &&   \delta_{n,0}  \left(\begin{array}{c} \delta_{s,1}\,\rho_{+,0}\lambda_+\\
    \delta_{s,2} \, \lambda_- \end{array}\right)+ 
    \left(\begin{array}{c}  a_n\\ \rho_{-,n} b_n\end{array}\right) \\ \label{s12down} &&=
  \frac{\sigma_0+r\sigma_x}{\sqrt{\cal T}}  \left(\begin{array}{c}\rho_{+,n-1} c_{n-1}\\d_{n-1}\end{array}\right).
\end{eqnarray}
Repeating the above exercise for $s=3,4$ then gives another pair of recurrence equations.  
From the upper components, we find
\begin{eqnarray}
 &&   \delta_{n,0}  \left(\begin{array}{c} \delta_{s,3}\,\rho_{+,0}\lambda_+\\
    \delta_{s,4} \,  \lambda_- \end{array}\right)+ 
    \left(\begin{array}{c} c_n\\\rho_{-,n} d_n\end{array}\right)  \nonumber\\ &&=
  \frac{\sigma_0-r\sigma_x}{\sqrt{\cal T}}  \left(\begin{array}{c}\rho_{+,n-1} a_{n-1}\\b_{n-1}\end{array}\right).\label{s34up}
\end{eqnarray}
while the lower component gives
\begin{eqnarray}\nonumber
 &&   \delta_{n,0}  \left(\begin{array}{c} \delta_{s,3}\,\lambda_+\\
    \delta_{s,4} \, \rho_{-,0} \lambda_- \end{array}\right)+ 
    \left(\begin{array}{c} \rho_{+,n} c_n\\d_n\end{array}\right) \\ \label{s34down} &&=
  \frac{\sigma_0-r\sigma_x}{\sqrt{\cal T}}  \left(\begin{array}{c} a_{n+1}\\\rho_{-,n+1}b_{n+1}\end{array}\right).
\end{eqnarray}
We observe that symmetry relations connect the recurrence relations for $s\in \{1,2\}$ 
with those for $s\in \{3,4\}$,
\begin{eqnarray} \nonumber
    s\in \{1,2\} &\leftrightarrow & s\in \{3,4\}, \\ \label{sym1}
\{ a,b,c,d\} &\leftrightarrow & \{ c,d,a,b\}, \\ 
 r &\leftrightarrow & -r.\nonumber
\end{eqnarray}
We therefore only need to study the relations for, say, $s\in \{1,2\}$, where the result
for $s\in \{3,4\}$ directly follows from Eq.~\eqref{sym1}.

We then focus on states incident from the left side ($s=1,2$).  By eliminating $(a_n,b_n)$
in Eqs.~\eqref{s12up} and \eqref{s12down}, we obtain a closed
recurrence relation for the scattering amplitudes $(c_n,d_n)$, 
\begin{eqnarray}\nonumber
 X_{n,n+1}(r) \left(\begin{array}{c} c\\d\end{array}\right)_{n+1} &=& 
Y_{n,n-1}(r)  \left(\begin{array}{c} c\\d\end{array}\right)_{n-1} + \\
&+& \label{finals12}
\delta_{n,0} \left(\begin{array}{c}\delta_{s,1}\, J_+ \\ \delta_{s,2} \, J_- \end{array}\right),
\end{eqnarray}
which are
needed below for computing the current in terms of outgoing states at $x>0$.
We use the source terms
\begin{equation}\label{source}
    J_{\alpha=\pm} = \alpha\left (\rho_{\alpha,0}^{-1} - \rho_{\alpha,0}^{} \right) \lambda_\alpha,
\end{equation}
and $2\times 2$ matrices in right-left mover space,
\begin{eqnarray}\nonumber
 X_{n,n+1}(r) &=& \frac{1}{\sqrt{{\cal T}}} \left( \begin{array}{cc} 
 \rho_{+,n}^{-1} & r \rho_{+,n}^{-1}\rho_{-,n+1}^{}  \\ r & \rho_{-,n+1} 
 \end{array}  \right)   ,\\ \label{MARmat}
  Y_{n,n-1}(r) &=& \frac{1}{\sqrt{{\cal T}}} \left( \begin{array}{cc} 
 \rho_{+,n-1} & r \\ r \rho_{+,n-1}^{}\rho_{-,n}^{-1}  & \rho_{-,n}^{-1} 
 \end{array}  \right)  .
\end{eqnarray}
They are connected by the relation
\begin{equation}
X_{n,n+1}^{-1}(r) = Y_{n+1,n}(-r).
\end{equation}
Equation \eqref{finals12} implies that only odd values of $n$ will contribute finite scattering
amplitudes $(c_n,d_n)$, i.e., $c_n=d_n=0$ for even $n$. 
By virtue of the symmetry relations \eqref{sym1},  the  
recurrence relation for $s=3,4$ can be inferred without additional calculations.

Taking into account all four incident-state types $s$, the time-dependent 
expectation value of the charge current follows from Eq.~\eqref{totalCPR} 
with $\nu_{\alpha=\pm}(E)$ in Eq.~\eqref{dosdef} as
\begin{eqnarray} \label{MARcur}
&&\qquad \qquad    I(t)= \sum_{{\rm even}\, m} e^{-imeV t} \, I_m , \\ \nonumber
&&I_m = \frac{e}{2h} \sum_{\alpha} \int dE\,n_F(E)\nu_\alpha(E) I_\alpha^{(m)}(r,E) + (r\to -r), 
\end{eqnarray}
where the terms with reflection amplitude $r\to -r$ describe incident states of type $s\in {3,4}$.
It is convenient to define $c_{+,n}=c_n$ and $d_{+,n}=d_n$ for the $s=1$ solution 
of the recurrences, and similarly $c_{-,n}$ and $d_{-,n}$ for $s=2$, i.e., the index $\alpha=\pm$ in $c_{\alpha,n}$ and $d_{\alpha,n}$ corresponds to
$s\in \{1,2\}$.  With this convention and using Eq.~\eqref{rhoan}, for even $m$, we find
 \begin{eqnarray}\nonumber
    I_\alpha^{(m)} (r,E) &=&  \sum_{{\rm odd}\,n} \Bigl [ 
    c_{\alpha,n}^\ast c^{}_{\alpha,n+m} \left(1+\rho_{+,n}^\ast \rho^{}_{+,n+m}\right) \\
    &-&  d_{\alpha,n}^\ast d^{}_{\alpha,n+m} \left(1+\rho_{-,n}^\ast \rho^{}_{-,n+m}\right)
    \Bigr], \label{iam}
 \end{eqnarray}
 where ``$\ast$'' indicates complex conjugation. The symmetry relation
\begin{equation}
    I_\alpha^{(-m)} (r,E) = \left( I_\alpha^{(m)} (r,E) \right)^\ast
\end{equation}
then implies $I_{-m}^{}=I_m^\ast$ for the current harmonics in Eq.~\eqref{MARcur}.

The DC current-voltage characteristics follows from the $m=0$ component,
\begin{equation}
   I(V) = \frac{e}{2h} \sum_\alpha\int dE\, n_F(E)\nu_\alpha(E)
   \label{MARcurDC}  I^{(0)}_\alpha(r,E)+(r\to -r).
\end{equation}
The current expression \eqref{MARcurDC}  affords a transparent physical interpretation.  Summing over all scattering channels $s$ and integrating over all incident energies $E$, the
weight of the corresponding incident state in the current is determined by the product of the Fermi function, the density of states, and a current matrix element.  
The latter follows by summing over all orders $n$ of the MAR ladder, where current contributions
only arise  for odd $n$.  At given order $n$,  electrons $(\propto |c_{\alpha,n}|^2)$ and holes $(\propto |d_{\alpha,n}|^2)$ enter with 
opposite sign, where the corresponding Doppler-shifted energy $E_n\mp v_Fq$ appears in the Andreev reflection amplitude.

\subsection{Ballistic limit} \label{sec5b}

Let us first discuss the ballistic case, ${\cal T}=1$, where the reflection amplitude vanishes ($r=0$) 
and a closed analytical solution is possible.

\subsubsection{Solution of the recurrence relations}

We first note that the matrices \eqref{MARmat} appearing in the recurrence relations \eqref{finals12}
are diagonal for $r=0$. 
For $s=1$, Eq.~\eqref{finals12} is then solved by $d_{+,n}=0$ (for all $n$), where
$c_{+,n}\ne 0$ is possible for odd $n$.  We thus obtain from Eq.~\eqref{finals12} the recurrence
relation
\begin{equation}\label{crec}
    c_{+,n+2}=\rho_{+,n+1}\rho_{+,n}\, c_{+,n} + \delta_{n,-1} \rho_{+,0} J_+ ,
\end{equation}
with $\rho_{\alpha,n}$ in Eq.~\eqref{rhoan} and $J_\alpha$ in Eq.~\eqref{source}.  
In order to obtain convergent solutions at large energies, 
we require $c_{+,-1}=0$.   As a  consequence, we find $c_{+,n}=0$
for all $n<0$.  Non-vanishing coefficients are then given by 
\begin{equation}\label{cps1}
    c_{+,n} = J_+\prod_{k=0}^{n-1} \rho_{+,k}  , \quad n\in \{1,3,\cdots\}.
\end{equation}

Similarly, for $s=2$, one finds $c_{-,n}=0$ for all $n$ while $d_{-,n}\ne 0$ is 
possible for odd $n$ according to the recurrence relation
\begin{equation}
    d_{-,n-2}=\rho_{-,n-1}\rho_{-,n} \, d_{-,n} - \delta_{n,1} \rho_{-,0} J_- ,
\end{equation}
where convergent solutions require $d_{-,1}=0$. We therefore must have $d_{-,n}=0$
for all $n>0$, and the non-vanishing coefficients are given by
\begin{equation}\label{dms2}
     d_{-,n} = -J_- \prod_{k=0}^{n+1} \rho_{-,k} , \quad n\in \{-1,-3,\cdots\}.
\end{equation}
Using the relation
$\rho_{\alpha,0} J_\alpha = \alpha \sqrt{2(1-\rho_{\alpha,0}^2)/\nu_\alpha(E)},$
Eqs.~\eqref{cps1} and \eqref{dms2} imply that all non-vanishing coefficients $c_{\alpha,n}$ 
and $d_{\alpha,n}$ for given energy $E$ can be summarized as follows.  
For electron-like quasiparticles (described by $c_{+,n}$), we find 
\begin{eqnarray}\nonumber
\frac{c_{+,n}}{c_{+,1}} &=& \prod_{k=1}^{n-1} \rho(E+keV-v_Fq) ,\quad (n=3,5,\cdots),\\
c_{+,1} &=& \sqrt{\frac{2}{\nu_+(E)}} \left[ 1-\rho^2(E-v_Fq)\right]^{1/2},\label{sol1}
\end{eqnarray}
with $\rho(E)$ in Eq.~\eqref{tildegamma}.
For hole-like quasiparticles ($d_{-,n}$), we find
\begin{eqnarray}\nonumber
\frac{d_{-,n}}{d_{-,-1}} &=& \prod_{k=-1}^{n+1} \rho(E+keV+v_Fq) ,\quad (n=-3,-5,\cdots),\\
d_{-,-1} &=& \sqrt{\frac{2}{\nu_-(E)}} \left[ 1-\rho^2(E+v_Fq)\right]^{1/2}.\label{sol2}
\end{eqnarray}
As a consequence, electron and hole type quasiparticles
climb the MAR ladder in opposite directions and their 
trajectories in energy space are shifted by $-v_Fq$ and $+v_Fq$, respectively.\\

\subsubsection{DC current-voltage curve}

The DC current-voltage curve then follows by inserting
Eqs.~\eqref{sol1} and \eqref{sol2} into Eq.~\eqref{MARcurDC}. 
Using $R(E)=|\rho(E)|^2$, we find 
\begin{widetext}
\begin{eqnarray}\label{marcur12}
I(V) &=& \frac{e}{h}\sum_\alpha \int dE\, n_F(E)\nu_\alpha(E) \sum_{{\rm odd}\,n} \left[ |c_{\alpha,n}|^2 \left(1+R(E+neV+v_Fq)\right) 
- |d_{\alpha,n}|^2 \left(1+R(E+neV-v_Fq)\right) \right] \\ \nonumber
&=& \frac{2e^2}{h} V + \frac{2e}{h}\sum_\alpha \int dE\, n_F(E) \Theta(|E-\alpha v_Fq|-\Delta) \,
\left[ 1-R(E-\alpha v_F q) \right] \left( 1+\sum_{m=1}^\infty \prod_{n=1}^m R(E+\alpha(neV-v_Fq))\right),
\end{eqnarray}
where we have carefully taken the large-bandwidth limit in the second step such that the Ohmic conductance due to direct quasiparticle tunneling processes (without Andreev reflections) is made explicit, see also Ref.~\cite{Hurd1997}.
For general transparency ${\cal T}$, the $I(V)$ curve contains the corresponding term $G_\infty V$ with $G_\infty$ in Eq.~\eqref{ohm}. 
In the zero-temperature limit, Eq.~\eqref{marcur12} can be simplified to 
\begin{equation}\label{ballisticMAR}
    I(V) = \frac{2e^2}{h} V -\frac{2e}{h}\int dE\, {\rm sgn}(E)\, \Theta\left(|E-v_Fq|-\Delta\right) 
\, [1-R(E-v_F q)] \, \left( 1+ \sum_{m=1}^\infty \prod_{n=1}^m R(E+neV-v_Fq)\right). 
\end{equation}
\end{widetext}
We observe that the current is determined by quasiparticle leakage into the leads due to MAR processes of all orders. 
The finite-$T$ result for $I(V)$ follows from Eq.~\eqref{ballisticMAR} by replacing ${\rm sgn}(E)\to \tanh[E/(2k_BT)]$.

From Eq.~\eqref{ballisticMAR}, we directly observe that $I(-V)=-I(V)$ for $q=0$. 
Rectification is possible, however, for $q\ne 0$.  Let us show this first for the limit $eV\gg \Delta$ with
$T=0$. Using $R(E)=\frac{\Delta^2}{4E^2}\ll 1$ for $E\gg \Delta$, see Eq.~\eqref{tildegamma},
we obtain from Eq.~\eqref{ballisticMAR}, up to terms of order $\Delta^2/(eV)^2$,
\begin{eqnarray} \nonumber
    I(V) &=& \frac{2e^2}{h}V-\frac{2e}{h}\int dE\, {\rm sgn}(E)\, \Theta\left(|E-v_Fq|-\Delta\right)\\
    &\times& \nonumber \left[ 1+R(E+eV-v_F q)\right]  
    \\ \nonumber
    &=&\frac{2e^2}{h}V +\frac{2e}{h} \lim_{\Omega\to \infty} \left(\int_{-\Omega}^{-\Delta+v_Fq} - \int_{\Delta+v_Fq}^\Omega\right) dE \times \\
    \nonumber &\times& \left[ 1+R(E+eV-v_F q)\right] \\
 \label{MARlargeV}  &=& \frac{2e}{h}\left[ eV+ 2( {\rm sgn}(V)+q\xi)\Delta \right]. 
\end{eqnarray}
In the ballistic limit, the current \eqref{MARlargeV} splits into the sum of the well-known current $I_{q=0}(V)$ for $q=0$ \cite{Averin1995} (here taken in the large-$V$ limit) plus a constant current $(v_F/h) (2e)(2q)$ carried by Cooper pairs propagating with finite momentum $2q$ and charge $2e$.  
A similar decomposition of the current is also found for the ${\cal T}=1$ equilibrium CPR, see Eq.~\eqref{Icontball}.

The above separation is in fact valid for arbitrary voltage $V$ as long as ${\cal T}=1$. Technically, one can show this by  shifting the integration variable $E\to E+v_Fq$ in Eq.~\eqref{ballisticMAR}.  For arbitrary $V$, we then obtain 
\begin{equation}\label{currdopplershift}
    I_{q\ne 0}(V) = I_{q=0}(V)+ \frac{4e\Delta}{h}  q\xi.  
\end{equation}

\subsubsection{Higher harmonics at low voltages}

Before turning to the general case with ${\cal T}<1$ in Sec.~\ref{sec5c}, 
let us address the higher harmonics of the AC current in the deep subgap 
regime $e|V|\ll \Delta$, still for ${\cal T}=1$ and $T=0$.  Here 
only incoming states with $E_\alpha <-\Delta$ (where $E_\alpha=E-\alpha v_Fq$ and $\alpha=\pm$) 
contribute to the current.  In order to simplify the analysis, we note that
the Andreev reflection amplitude $\rho(E)$ in Eq.~\eqref{tildegamma} is very small for above-gap
energies $E>\Delta$, and we therefore assume $\rho(E>\Delta)=0$ below.

We next observe that for scattering channel $s=1$, where states with incident energy $E$
are described by the scattering amplitudes $c_{+,n}$ with $n\in \{1,3,5,\ldots\}$, contributions
to the MAR ladder are only possible for the energy window $-\Delta-eV<E-v_Fq<-\Delta$. 
Indeed, only for such energies, one can have steps with $\rho(E+keV-v_Fq)=1$ in Eq.~\eqref{sol1}.
Note that this window requires $V>0$.
Similarly, for $s=2$, the amplitudes $d_{-,n}$ with $n\in\{-1,-3,\ldots\}$ in Eq.~\eqref{sol2}
imply that only states with $-\Delta+eV<E+v_Fq<-\Delta$ contribute 
to the MAR ladder and we must have $V<0$.
As a consequence,  for $V>0$ $(V<0)$, only states of incoming type $s=1$ ($s=2$)
can climb the MAR ladder and thus contribute to the current.  They
must originate from a narrow energy region near the gap edges, $-\Delta-e|V|< E_\alpha< -\Delta$.

Let us then consider the case $0<eV\ll \Delta$, where within our approximations, we have $d_{-,n}=0$ from Eq.~\eqref{sol2} and $I_-^{(m)}=0$ from Eq.~\eqref{iam}. 
Truncating the MAR ladder at order $n_{\rm max}\approx 2\Delta/eV \gg 1$, the amplitudes
in Eq.~\eqref{sol1} are now given by
\begin{eqnarray}\nonumber
c_{+,n} &=& e^{-i\sum_{k=1}^{n-1} \gamma_{+,k}}\, c_{+,1} ,\qquad c_{+,1} = \sqrt{\frac{2}{\nu_+(E)}},\\
\gamma_{\alpha,k} &=& \cos^{-1}\left(\frac{E+keV-\alpha v_Fq}{\Delta}\right)\in [0,\pi],
\label{sol1lowv}
\end{eqnarray}
where $n=3,5,\cdots,n_{\rm max}$. The non-vanishing terms
 in Eq.~\eqref{iam} have the form (with $r=0$ and even $m$)
\begin{equation}
    I_+^{(m)}(0,E) = \sum_{n=1}^{n_{\rm max}-m} c_{+,n}^\ast c_{+,n+m}^{} \left(1+ 
    e^{i(\gamma_{+,n}-\gamma_{+,n+m})}\right). 
\end{equation}
The current harmonics $I_m$ for $m=2,4,\ldots$, see Eq.~\eqref{MARcur}, 
are then given by
\begin{eqnarray} \nonumber
I_m &=& \frac{e}{h} \int_{-\infty}^0 dE \, \nu_+(E) I_+^{(m)}(0,E) \\ \nonumber
&\simeq & \frac{2e}{h} \int_{-\Delta+v_Fq -eV}^{-\Delta+v_Fq} dE \sum_{k=1}^{n_{\rm max}-m}
e^{-i\sum_{j=0}^{m-1} \gamma_{+,k+j}} \\
&=& \frac{2e}{h} \int_{-\Delta -eV}^{-\Delta} dE \sum_{k=1}^{n_{\rm max}-m} e^{-i\sum_{j=0}^{m-1} \gamma_{k+j}},
\end{eqnarray}
where $\gamma_k\equiv \gamma_{+,k}(q=0)$ and we have shifted the energy integration variable in the 
last step. We thus arrive at the $q=0$ result of Ref.~\cite{Averin1995},
\begin{equation}
    I_m = \frac{2\Delta}{h} \int_{-1}^1 dz \, e^{-im \cos^{-1}z},
\end{equation}
which continues to hold for $q\ne 0$ and $m\ne 0$.
We conclude that the higher current harmonics are \emph{not} affected by the $q$-shift
of the DC current in Eq.~\eqref{currdopplershift}.
We emphasize that the $q$-independence of $I_{m\ne 0}$ 
holds in the ballistic limit and for low voltages, where the
above derivation is justified.  It is an open question if and how the current harmonics
are affected by $q\ne 0$ otherwise.

Neglecting the (now subleading) Ohmic contribution, we conclude that 
 for $e|V|\ll \Delta$ and $T=0$, the time-dependent current
$I(t)$ in Eq.~\eqref{MARcur} is given by
\begin{equation}\label{fulllowv}
    I(t) = \frac{4e\Delta}{h}  q\xi + \frac{e\Delta}{\hbar} {\rm sgn}(V)\, \left|\sin(eVt)\right|. 
\end{equation}
The time-averaged current obtained from Eq.~\eqref{fulllowv} is consistent with the DC current in  Eq.~\eqref{currdopplershift} since $I_0\simeq (4e\Delta/h)\, {\rm sgn}(V)$ for $e|V|\ll \Delta$ \cite{Averin1995}.
We then obtain for $e|V|\ll \Delta$ the DC current-voltage curve as
\begin{equation}\label{ivf}
    I(V) = \frac{4e\Delta}{h} \left( q\xi+{\rm sgn}(V)\right).
\end{equation}
For $q\xi\to 1$, we obtain $I(V>0)=8e\Delta/h$ and $I(V<0)=0$, resulting in full rectification with 
maximal efficiency $\eta(V)=1$, see Sec.~\ref{sec6}.

In addition, Eq.~\eqref{fulllowv} is also consistent with our results for the CPR of the ballistic
Josephson diode, see Sec.~\ref{sec3b}, with two Andreev states $\pm E_A(\varphi)$ for
$E_A(\varphi)=\Delta\cos(\varphi/2)-v_Fq$.  At small but finite voltage $V$, these Andreev level 
energies acquire an adiabatic time dependence because of the Josephson relation $\varphi(t)=2eV t$.  
For instance, starting from $E_A=-\Delta$ at time $t=0$, we have $E_A(t)=\Delta \cos[eV(t-t_0)]$ with $eVt_0=\cos^{-1}(q\xi-1)$.  In the ballistic limit with $q\ne 0$, and assuming that no other dissipation channels are
present, each Andreev level for chirality $\alpha$ may dive into (or out of) the continuum associated with the opposite chirality $-\alpha$ without changing its occupation probability.  The reason for this 
behavior has been discussed in Sec.~\ref{sec2b}: even though there is a spectral overlap of states
with $\alpha=+$ and $\alpha=-$, and therefore between continuum and Andreev states, both types of states
are completely decoupled in the ballistic case.  As a consequence, the occupation probability 
of Andreev levels can only change near gap edges of the same chirality.  

\subsection{Non-ideal transparency} \label{sec5c}

\begin{figure}
\centering
 \includegraphics[width=0.48\textwidth]{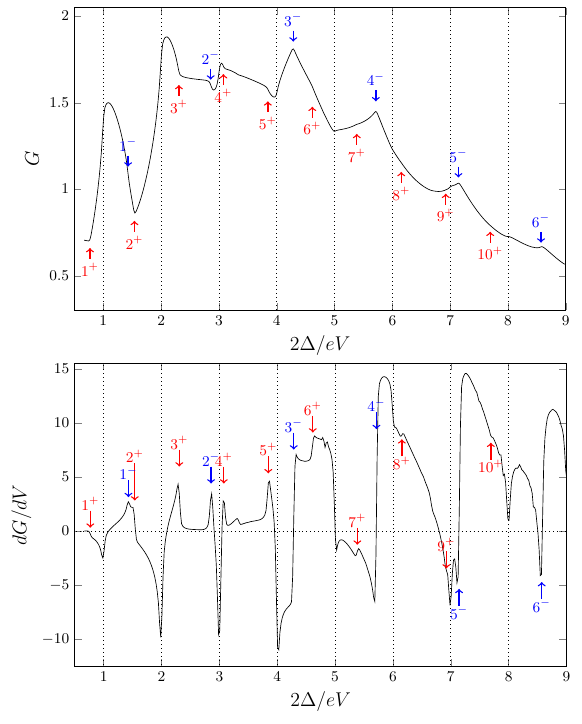}
\includegraphics[width=0.48\textwidth]{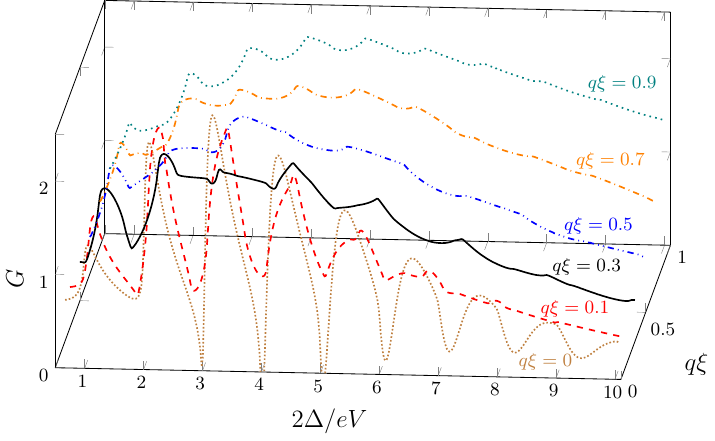}
\caption{Differential conductance $G=dI/dV$ (top and center panel, in units of $G_0=2e^2/h$),
and derivative $dG/dV$ (bottom panel, in units of $eG_0/\Delta$) 
 of a Josephson diode as function of the voltage parameter
$2\Delta/eV$  for $T=10^{-3}\Delta$, ${\cal T}=0.7$ and $q\xi=0.3$.  
The bottom panel shows a waterfall plot for ${\cal T}=0.7$ and several $q\xi$.  
Red and blue arrows labeled by $n^\pm$ correspond to MAR features at $2\Delta_\pm/eV=n$, respectively, with $\Delta_\pm$ in Eq.~\eqref{Dpm}.  MAR features at $2\Delta/eV=n$ correspond to the dotted vertical lines.  }
\label{fig5} 
\end{figure}

We now show results for the $I$-$V$ characteristics for voltage-biased Josephson diodes with finite CP momentum $2q\ne 0$ and transparency ${\cal T}<1$. 
We focus on the case $T=0$.  The results shown below have been obtained from Eq.~\eqref{MARcurDC} 
by numerical solution of the recursion relation \eqref{finals12}.  
It is a well-known challenge to describe the limit $V\to 0$ with ${\cal T}\alt 1$ numerically \cite{Bratus1995,Averin1995,Cuevas1996,Zazunov2006} 
since the MAR order contributing to the current, $n_{\rm max}\sim 2\Delta/e|V|$, becomes very large. As a check, our code reproduces (i) the $q=0$ results of Ref.~\cite{Averin1995} and (ii) the ballistic result for $q\ne 0$ in Eq.~\eqref{currdopplershift}.

In Fig.~\ref{fig5}, we illustrate our numerical results for the differential conductance $G(V)=dI/dV$ and for the derivative $dG/dV=d^2I/dV^2$. We find a rich subharmonic structure, with sharp peaks and dips at certain voltages.
In particular, we find the conventional MAR features at voltages $eV_n=2\Delta/n$ (integer $n$) but also shifted features at $2\Delta_\pm/eV=n$ corresponding to the Doppler shifted spectral gaps in Eq.~\eqref{Dpm}.   

An example for a typical MAR trajectory (representing the MAR ladder in energy space)
in the ballistic limit is schematically illustrated in the top panel of Fig.~\ref{fig4}.  
Because of the characteristic peak structure near the band edge in the superconducting density of states \eqref{dosdef}, the shown process in Fig.~\ref{fig4} with an incoming electron of energy $E\approx -\Delta_-$ will have a resonant enhancement if the energy $E+3eV$ of the outgoing electron is just above $\Delta_+$. The MAR resonance then occurs if $2\Delta/eV=3$ since $\Delta_-+\Delta_+=2\Delta$, see also Fig.~\ref{fig1}(b).  
More generally, similar MAR trajectories can explain the occurrence of
MAR features at voltages where $2\Delta/eV=n$ with integer $n$.  

In the absence of normal reflection, MAR trajectories such as the one shown in the top panel of Fig.~\ref{fig4} 
cannot end in an outgoing electron state with  $E_n\approx +\Delta_-$ since the corresponding 
matrix elements vanish.  However, for finite reflection amplitude $r$, i.e., for ${\cal T}<1$, 
MAR trajectories are able to connect quasiparticle states with incident energy $E\approx -\Delta_\pm$ and outgoing energy
$E_n\approx +\Delta_\pm$.  In such cases, depending on the magnitude of the current matrix elements,
one can obtain MAR side features at $eV=2\Delta_\pm/n$.
An example for such a case is shown in the lower panel of Fig.~\ref{fig4}.

Such features are clearly observed from our numerical results in Fig.~\ref{fig5}, where
side peaks or dips for $2\Delta_\pm/eV=n$ are
indicated by red and blue arrows, respectively.
While some of those features are hardly visible in the nonlinear conductance (top panel), the 
derivative $dG/dV$ (center panel) reveals sharp features.
We  remark that for some points with $2\Delta_\pm/eV=n$ in Fig.~\ref{fig5}, MAR features are
(almost) absent since the corresponding current matrix elements are very small.
This happens, for instance, near $2\Delta_-/eV\approx 8$.  

We conclude that the rich subharmonic structure in Fig.~\ref{fig5} is connected to the 
pair of Doppler shifted spectral gaps.  The presence of two gaps enriches the MAR ladder 
picture and implies that MAR features can occur not only for $2\Delta/eV=n$ but also for
$2\Delta_\pm/eV=n$.

\section{Finite-voltage rectification} \label{sec6}

In this section, we discuss the finite-bias rectification efficiency in the zero-temperature limit.
For arbitrary system parameters represented by the dimensionless quantities $q\xi$, ${\cal T}$, and $eV/\Delta$,
the  efficiency $\eta(V)$ in Eq.~\eqref{efficiency} follows from Eq.~\eqref{MARcur} 
by numerically solving the recurrence relations.  In Sec.~\ref{sec6a}, we show that the perfect efficiency
$\eta=1$ is reachable in the ballistic case ${\cal T}=1$.  We then turn to the subharmonic structure for
${\cal T}<1$ in Sec.~\ref{sec6b}, and finally comment on the large-bias regime in Sec.~\ref{sec6c}.

 \subsection{Ideal rectification}\label{sec6a}

For ${\cal T}=1$, the matching conditions as well as the BdG Hamiltonian conserve chirality, $\sigma_z=\alpha=\pm$, and the recurrence relations admit a closed solution.
As discussed in Sec.~\ref{sec5b}, we find that $I(V)$ for $q\ne 0$ is related to the known $q=0$ curve $I_{q=0}(V)$ \cite{Averin1995}  by a simple shift, see Eq.~\eqref{currdopplershift}.
This shift has a clear physical interpretation: it is the current carried by Cooper pairs with finite momentum $2q$ and charge $2e$.  The decomposition \eqref{currdopplershift} only 
applies in the ballistic limit where chirality is conserved.   
The rectification efficiency  then follows from Eq.~\eqref{efficiency} as
\begin{equation}\label{rectball}
    \eta(V,q\xi,{\cal T}=1) = \frac{4e\Delta}{h} \frac{q\xi}{I_{q=0}(V)},
\end{equation}
which depends on the voltage only through the ratio $eV/\Delta$, i.e., 
the Doppler shifted gaps $\Delta_\pm$ do not appear.
For $e|V|\gg \Delta$, the Ohmic result of a normal-conducting contact, $I_0(V)\approx (2e^2/h)V$, implies $\eta(V) \simeq  2q\xi \frac{\Delta}{eV}$.
On the other hand, for $e|V|\ll \Delta$, using $I_{q=0}(V\to 0)\approx (4e \Delta/h)\,{\rm sgn}(V)$ \cite{Averin1995}, Eq.~\eqref{rectball} gives $\eta(V)\simeq q\xi.$  Remarkably,
for $q\xi\to 1$, one thus approaches the \emph{ideal rectification limit}  since the MAR-induced current $I_0$ now precisely cancels the finite-momentum Cooper pair current for $V<0$, i.e, $I(V<0)=0$ in Eq.~\eqref{currdopplershift}, while both currents add for $V>0$ to give $I(V>0)=8e\Delta/h$. 
 As a result, we have $\eta(V)=1$. Clearly, there is no current suppression for $q\xi\to 1$ even though one of the spectral gaps vanishes, $\Delta_-\to 0$.
 
We conclude that MAR processes can generate highly efficient superconducting diode 
behavior along with large currents in the deep subgap regime $e|V|\ll \Delta$ and at low temperatures.
The perfect rectification limit is reached for ballistic junctions with CP momentum parameter $q\xi\to 1$.

\subsection{Subharmonic structure}\label{sec6b}

\begin{figure}
\includegraphics[width=0.49\textwidth]{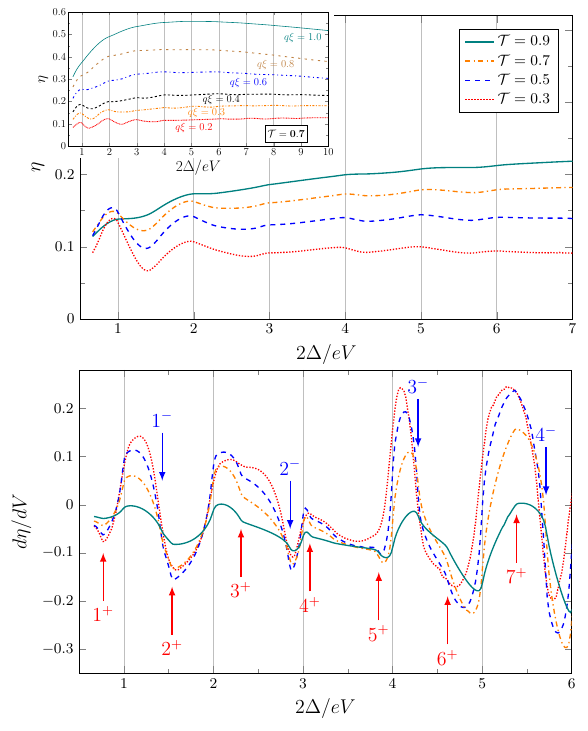}
\caption{Rectification efficiency $\eta(V)$ obtained by numerical evaluation of Eqs.~\eqref{efficiency} and \eqref{MARcur}. 
Top panel: The main part shows $\eta(V)$ vs $2\Delta/eV$ 
for $q\xi=0.3$ and several ${\cal T}$.  Dotted vertical lines indicate standard 
MAR features at $2\Delta/eV=n$ (integer $n$) \cite{Klapwijk1982,Bratus1995,Averin1995,Cuevas1996}.
The inset is for ${\cal T}=0.7$ and different $q\xi$. 
Bottom panel:  $d\eta/dV$ vs $2\Delta/eV$ for $q\xi=0.3$ and the same ${\cal T}$ as in the 
main part of the top panel.
Arrows labeled by $n^\pm$ indicate the points where $2\Delta_\pm/eV=n$. 
}
\label{fig6}
\end{figure}

\begin{figure}
\includegraphics[width=0.47\textwidth]{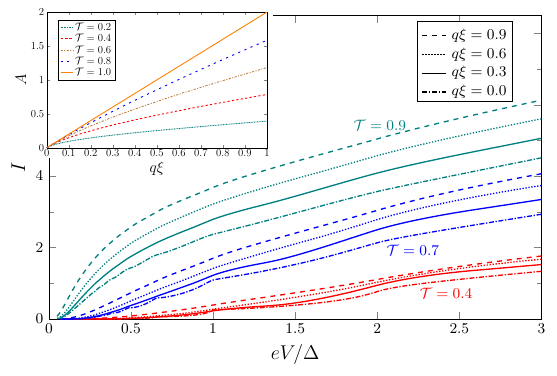}
\caption{$I$-$V$ curves and high-voltage rectification. 
Main panel:  $I$-$V$ curves for different values of $(q\xi,{\cal T})$, with currents in units of $2e\Delta/h$.  Inset: Rectification coefficient $A$ vs $q\xi$ for several ${\cal T}$ in the large-voltage limit, see Eq.~\eqref{Adef}, obtained by numerical solution
of Eq.~\eqref{MARcur} for $eV/\Delta=50$.
 }
\label{fig7}
\end{figure}

In Fig.~\ref{fig6}, we show numerical results for 
$\eta(V)$ for different values of $(q\xi,{\cal T})$.  
We observe an overall increase of $\eta(V)$ with increasing Cooper pair 
momentum $2q$ and/or junction transparency ${\cal T}$.
The efficiency is particularly large in the subgap regime $e |V|\alt 2\Delta$, where we also observe a subharmonic structure with peaks or dips. 
The enhancement of $\eta(V)$ for $e|V|\ll \Delta$  compared to $\eta_0$ 
(for otherwise identical parameters) thus persists for ${\cal T}<1$, 
with optimal rectification at ${\cal T}=1$.
The subharmonic structure is more clearly visible 
in the derivative $d\eta(V)/dV$ (bottom panel in Fig.~\ref{fig6}).
Apart from the standard $q=0$ MAR features at $2\Delta/eV=n$ (integer $n$),
which are also observed for $q\ne 0$ and follow from MAR trajectories as 
drawn in the upper panel in Fig.~\ref{fig4}, we also find 
resonances or antiresonances corresponding to the Doppler-shifted pairing gaps $\Delta_\pm$ (indicated by arrows in Fig.~\ref{fig6}).  The physics of these features has already been described above for the nonlinear conductance:
The transitions are naturally explained from the MAR ladder picture, see the lower panel of Fig.~\ref{fig4}, 
where the presence of normal reflection $r\ne 0$ enables
MAR trajectories between states near the same type of spectral gap ($\pm\Delta_+$ or $\pm\Delta_-$) where $\nu_\alpha(E)$ has sharp peaks. 

\subsection{Large-bias behavior}\label{sec6c}

In Fig.~\ref{fig7}, we illustrate the $I$-$V$ curves for several
values of $(q\xi,{\cal T})$. Note that the current remains large for $q\xi\to 1$, indicating
efficient diode behavior along with large currents.
For $eV\gg \Delta$, we find
\begin{equation}\label{Adef}
\eta(eV\gg\Delta,q\xi,{\cal T}) \simeq A(q\xi,{\cal T}) \frac{\Delta}{eV}.
\end{equation}
The dimensionless coefficient $A(q\xi,{\cal T})$ is shown
in the inset of Fig.~\ref{fig7}, with $A=2q\xi$ for ${\cal T}=1$ from the analytical solution. 
For $q\xi \alt 1$, our numerical results suggest $A(q\xi,{\cal T})\approx 2q\xi{\cal T}$.

We conclude that even for large bias voltages well above the superconducting pairing gap, $eV\gg\Delta$, some degree of rectification can 
persist at low temperatures.  

%%%%%%%%%%%%%%%%%%%%%%%%%%%%%%%%%%%%%%%%%%%%%%%%%%%%%%%%%%%%%%%%%%%%%%%%%%%%%%%%%%%%%%%%%%%%%
\section{Conclusions and outlook} \label{sec7}

In this work, we have studied the physics of superconducting diodes when operated under voltage-biased nonequilibrium conditions.  
By developing a scattering approach for a single-channel junction involving helical superconductors with finite CP momentum $2q$, 
the current-voltage characteristics can be computed, fully including MAR processes to all orders. 
Apart from the case of voltage-biased Josephson junctions, we have also studied the equilibrium case and the case of NS junctions. Our main results are as follows.  

First, for the equilibrium case corresponding to a current-biased junction, the
SDE efficiency $\eta_0$ reaches a maximal value of $\eta_0\approx 0.4$ in the limit of
a ballistic junction (transparency ${\cal T}=1$) with $T=0$ and CP momentum parameter $q\xi\approx 0.9$. These results  reproduce those in Refs.~\cite{Davydova2022,FuErratum}.

Second, for the NS junction, the nonlinear $T=0$ conductance $G(V)=dI/dV$ exhibits discontinuities in the derivative $dG/dV$ at the voltages corresponding to  Doppler shifted spectral gaps, $eV=\Delta_\pm=\Delta\pm v_Fq$.
Tunneling spectroscopy of helical superconductors is thus predicted to find clear traces of the finite CP momentum through the doubling of such singular features. However, the NS junction can never exhibit rectification.

Third, for voltage-biased Josephson diodes, we have developed a scattering approach including MAR processes to all orders.  Our theory reproduces the results of 
Ref.~\cite{Averin1995} in the limit of vanishing CP momentum.
For the ballistic case, an analytical solution has been presented which facilitates the
calculation of the current-voltage characteristics.  We can thereby compute the rectification efficiency $\eta(V)$ as a function of the dimensionless parameters of this problem (which are 
given by $q\xi,{\cal T},k_B T/\Delta$, and $eV/\Delta)$. 
We obtain large efficiencies approaching the ideal value $\eta(V)=1$ at low temperature and very small voltage, 
assuming a junction with
high transparency ${\cal T}\to 1$ and CP parameter $q\xi\to 1$. 
While the SDE efficiency $\eta_0$ is always smaller than
$\eta_0\approx 0.4$, going out of equilibrium can thus result in ideal rectification.

Fourth, for finite CP momentum, we predict that the standard MAR features 
at $2\Delta/eV=n$ (integer $n$) will be accompanied by side features (peaks or dips) at  $2\Delta_\pm/eV=n$ determined by the Doppler shifted spectral gaps $\Delta_\pm$.

A central and remarkable result of our work is that in the subgap regime, MAR processes can allow for very large rectification efficiency, accompanied by large currents. 
The $I$-$V$ curve for $e|V|\ll \Delta$ can be computed analytically  
from a time average over quasi-stationary Andreev levels \cite{Averin1995} by using the Josephson relation $\dot\varphi=2eV/\hbar$.  In this case, relaxation mechanisms mixing Andreev states are inefficient and one can obtain perfect rectification ($\eta=1$) at 
the optimal working point $q\xi=1$. On the other hand, under current-biased conditions, an effective relaxation mechanism is tacitly assumed, and this results in  $\eta_0\alt 0.4$ and a different optimal working point ($q\xi\approx 0.9$). 
One may expect a similar rectification efficiency enhancement in voltage-biased Josephson diodes
also if other mechanisms are responsible for the SDE. 
We hope that future experimental and theoretical work will shed light on this intriguing question.

Apart from obvious implications for experiments on Josephson diode, our work raises several 
interesting questions for future theoretical research.  Let us mention just a few points. 
First, a relatively straightforward extension
of our theory is to analyze the higher current harmonics $I_{m\ne 0}$ in Eq.~\eqref{MARcur}.
While we have studied them in the ballistic limit at
very low voltages, the case of ${\cal T}<1$ and larger voltages remains open.
Second, it would be of interest to  analyze the gapless
case realized for large CP momentum with $q\xi>1$, which has not been 
discussed in the present work.  Third, in the DC limit, quantum noise could exhibit 
interesting features in a voltage-biased Josephson junction with $q\ne 0$. 
Fourth, the above MAR recurrence relations may also allow 
for analytical progress by expanding in the reflection amplitude $r=\sqrt{1-{\cal T}}$ 
for $r\ll 1$, see Ref.~\cite{Zaikin2023} for recent work on the $q=0$ version of the model.

Finally, MAR-related effects could be different in cases where the mechanism behind
 the SDE is not based on finite CP momentum physics.  In App.~\ref{appA}, we briefly discuss a model with $q=0$, where the weak link is defined by a tunnel-coupled quantum dot with spin-orbit coupling and a magnetic Zeeman field.  This model exhibits the SDE and, according to our preliminary results, the efficiency $\eta(V)$ can again be large in the low-voltage regime at high transparency. Such models in addition allow one to study Coulomb interaction effects.  

 To conclude, we hope that our work will inspire future research on nonequilibrium transport in systems exhibiting the SDE.

\begin{acknowledgments} 
We thank Liang Fu for discussions.
We acknowledge funding by the Deutsche Forschungsgemeinschaft (DFG, German Research Foundation) under Grant No.~277101999 - TRR 183 (project C01), Grant No.~EG 96/13-1, and under Germany's Excellence Strategy - Cluster of Excellence Matter and Light for Quantum Computing (ML4Q) EXC 2004/1 - 390534769. 
This work received support from the French government under the France 2030 investment plan, as part of the Initiative d'Excellence d'Aix-Marseille Universit\'e - A*MIDEX, through the institutes IPhU (AMX-19-IET-008) and AMUtech (AMX-19-IET-01X).
\end{acknowledgments}

%%%%%%%%%%%%%%%%%%%%%%%%%%%%%%%%%%%%%%%%%%%%%%%%%%%%%%%%%%%%%%%%%%%%%%%%%%%%%%%%%%%%%%%%%%%%%%%%%%%%%%
\appendix
\section{Quantum dot Josephson diode model}\label{appA}

\begin{figure}
   \centering
    \includegraphics[width=0.49 \textwidth]{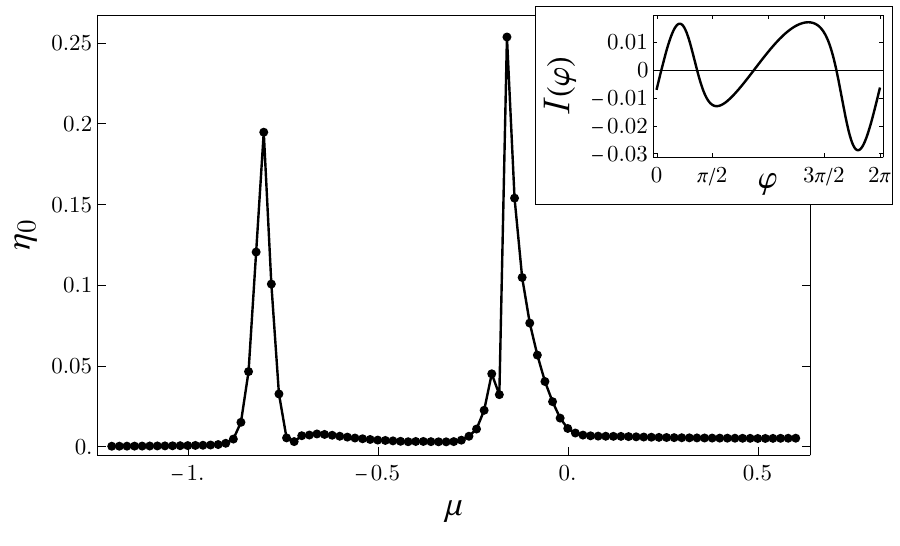}
  \caption{ SDE efficiency $\eta_0$ for the quantum dot model of 
  a Josephson diode in App.~\ref{appA}. We use units with $\Delta=1$ and 
  the parameters $\alpha_0=1.4$, $\mathbf{b}=(0.4,0,0.3)$, $L=3$, $t_L=0.45$, $t_R=0.4$, 
 and $k_B T=0.01$.  Main panel: $\eta_0$ vs dot potential $\mu$. 
 Lines connecting data points are a guide to the eye only.  Inset: CPR for $\mu=-0.16$, 
 where the SDE efficiency is $\eta_0 \approx 0.25$. }
  \label{fig8}
\end{figure}

\begin{figure}
\begin{tabular}{cc}
\adjustbox{valign=M}{\includegraphics[width=0.41 \textwidth]{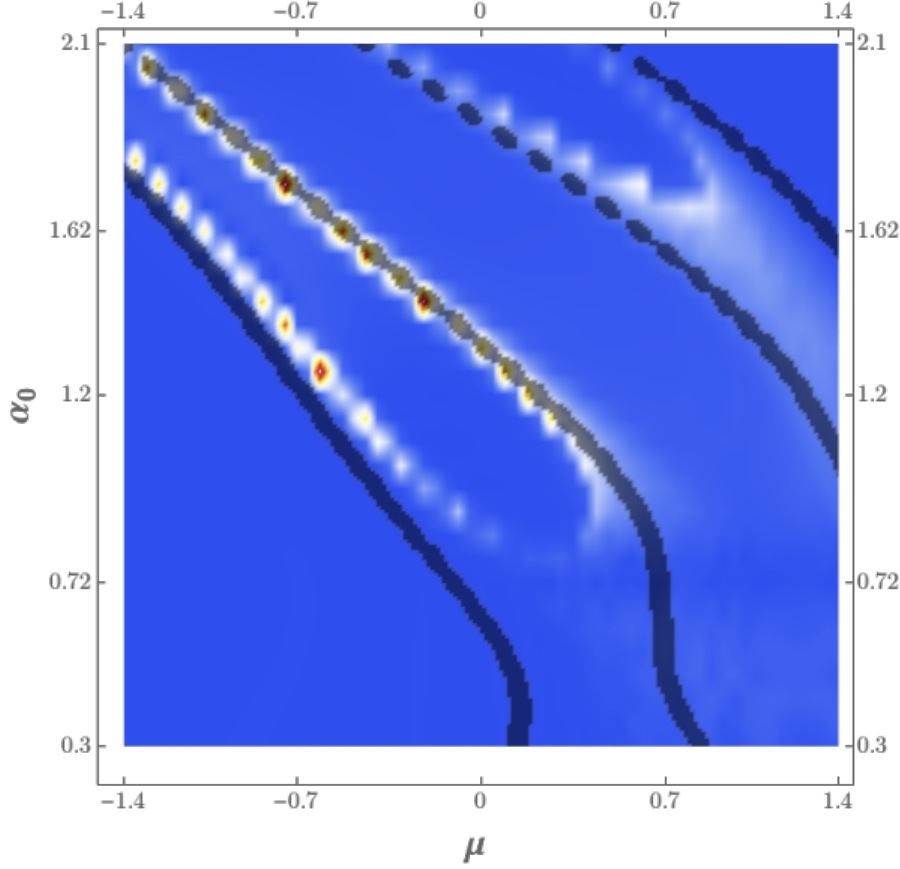}} &
\adjustbox{valign=M}{\includegraphics[width=0.06 \textwidth]{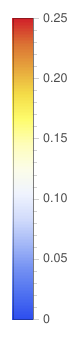}}
\end{tabular} 
\caption{ Two-dimensional color-scale plot of the SDE efficiency $\eta_0$, Eq.~\eqref{eta0},
in the $\alpha_0$-$\mu$ plane, where $\alpha_0$ is the spin-orbit coupling parameter and $\mu$ denotes the dot potential.
We use the parameters in Fig.~\ref{fig8} and units with $\Delta=1$.   The magnitude of $\eta_0$ is encoded by the color bar.
Superimposed thick black lines indicate positions in the $\alpha_0$-$\mu$ plane 
where a dot energy level $\epsilon$ is nearly resonant with $|\epsilon| < 0.04 \Delta$.  Regions with large $\eta_0$ typically are close to these resonance lines.}
\label{fig9}
\end{figure}

In this Appendix, we summarize our results for
the SDE efficiency $\eta_0$ in Eq.~\eqref{eta0} for a 
different Josephson diode model 
\cite{DellAnna2007,Zazunov2009,Brunetti2013}. In this model, the superconducting bands 
have zero CP momentum, $q=0$, but the weak link is defined by 
a quantum dot tunnel-coupled to the superconducting leads, 
$H = H_{L}+H_R + H_{\rm dot} + H_{\rm{tun}}$.
The dot is modeled as a nanowire of length $L$ along the $x$-axis with spin-orbit
and Zeeman couplings \cite{Zazunov2009}, 
\begin{equation}
H_{\rm dot}= \int_{-L/2}^{L/2} dx \; d^{\dagger}(x) 
\left(  \frac{\hat{p}^2}{2m} - \mu + \frac{\alpha_0}{m} \sigma_z \hat{p} 
+ \mathbf{b}\cdot \mathbf{\sigma}\right) d(x),
\label{eq:SOI_Hdot}
\end{equation}
where $d = (d_{\uparrow}, d_{\downarrow} )^T$ is a fermionic spinor field, 
$\hat{p} = -i \partial_x$, $m$ is the effective electron mass, $\mu$ is the chemical
potential of the dot, $\alpha_0 \geq 0$ is the spin-orbit coupling, and
$\mathbf{b} = (b_x , b_y , b_z)$ is a constant magnetic Zeeman field
(including gyromagnetic and Bohr magneton factors). In contrast to the notation 
used in the main text, the Pauli matrices $\sigma_{x,y,z}$ in Eq.~\eqref{eq:SOI_Hdot}
act in spin space. 

As in Eq.~\eqref{HLR}, the superconducting leads are described by 
\begin{equation}
    H_{j=L,R} =  \sum_{k} \psi^{\dagger}_{j,k}   \left( \xi_k \tau_z + \Delta \tau_x \right) \psi_{j,k}^{},
\label{eq:SOI_Hleads}
\end{equation}
with the Nambu spinor $\psi_{j,k} = (\psi_{j,k \uparrow}^{}, \psi^{\dagger}_{j, k \downarrow})^T$. Here $\psi_{j,k \sigma}^\dagger$ creates an electron in lead $j = L/R$ with momentum $k$, spin projection $\sigma \in \{ \uparrow, \downarrow\}$, and normal-state dispersion $\xi_k$.
The Pauli matrices $\tau_{x,y,z}$ act in Nambu space.

Finally, $H_{\rm tun}$ describes point-like tunnel couplings connecting the leads and the dot,
\begin{equation}
 H_{\rm tun} =  \sum_{j=L,R} t_j e^{i \phi_j/2} \sum_{\sigma=\uparrow,\downarrow} \psi^{\dagger}_{j \sigma} d_{\sigma}(x_j) + \text{H.c.}, 
 \label{eq:SOI_Htun}
\end{equation}
with momentum- and spin-independent real-valued hopping parameters $t_j > 0$. The
tunneling points in the nanowire are given by $x_j = -s_j L/2$ with $s_{L/R} = \pm 1$, and the  local lead fermion fields near the respective tunnel contact are $\psi_{j,\sigma} = 
 \sum_{k} \psi_{j,k \sigma}$. We again choose a gauge where $\phi_j = s_j \varphi/2$ with the phase difference $\varphi$.

Integrating out the superconducting lead fermions, we obtain the CPR 
in terms of the dot Green's function, see Refs.~\cite{DellAnna2007,Zazunov2009} for details. 
The resulting CPR depends on the various model parameters. However, 
from numerical calculations for many different parameter configurations,
the generic behavior is as follows, see also Ref.~\cite{Zazunov2009}.  
If both the spin-orbit parameter $\alpha_0$ and the magnetic Zeeman field $\mathbf{b}$ 
are finite, and if more than one dot level contributes to the Josephson current, 
the SDE efficiency $\eta_0$ in Eq.~\eqref{eta0} is generally finite. Importantly,
there are always specific regions in parameter space 
where the CPR implies a large SDE efficiency, $\eta_0\agt 0.2$. 
We here illustrate this conclusion for the case of relatively short nanowires, where precisely two
dot energy levels $\epsilon_n$ satisfy $|\epsilon_n|\alt \Delta$. 
(However, for very short wires, only a single dot level is present, but then the spin-orbit coupling is not effective and there is no SDE \cite{Zazunov2009}.)
In the case discussed below, all other dot levels have energy $|\epsilon|\gg \Delta$, and their contribution to the Josephson current is negligible.  
The dot energy levels $\epsilon_n$ follow by diagonalization
of Eq.~\eqref{eq:SOI_Hdot} \cite{DellAnna2007,Zazunov2009}.

Figure \ref{fig8} shows the resulting SDE parameter $\eta_0$
as a function of the dot potential $\mu$ for a generic parameter choice. 
We find that $\eta_0$ is very small for most values of 
$\mu$, except near two specific values where efficiency peaks
are present. The maximal SDE efficiency is here given by $\eta_0\approx 0.25$. 
The inset in Fig.~\ref{fig8} shows the CPR
for $\mu=-0.16$, corresponding to the point of maximal SDE efficiency in the main panel. The negative critical current is here much larger (in absolute value) than the positive critical current. Note that the CPR is strongly anharmonic, which is a necessary condition to obtain large $\eta_0$. We stress that the results shown
in Fig.~\ref{fig8} are generic:  similar results have been 
found for many different parameter choices.

Since a computation of $\eta_0$ requires knowledge of the full CPR, it is generally
hard to reliably predict the precise parameter values allowing for large efficiencies.
However, physical insight follows by scanning the positions in  
parameter space where a large $\eta_0$ occurs. 
In Fig.~\ref{fig9}, we show how $\eta_0$ depends on the spin-orbit coupling $\alpha_0$ 
and on the dot potential $\mu$, with the remaining parameters chosen 
as in Fig.~\ref{fig8}. In this two-dimensional plane, the points with large 
$\eta_0$ form a set of curves. For a large part, these curves coincide (or are very close)
with parameter values where one of the dot levels has energy close to zero. 
Such resonant dot energy positions are shown by thick black lines in Fig.~\ref{fig9}. 
Large values of $\eta_0$ are therefore tightly correlated with the existence of a (nearly) resonant dot level.
One can rationalize this conclusion by noting that resonant dot energy levels 
typically cause large transparency of the corresponding weak link, which in turn
can cause the strongly anharmonic CPR \cite{Nazarov2009} needed for 
generating a large SDE efficiency.  On a qualitative level, this conclusion is 
also consistent with the results for the finite CP momentum model presented in
Sec.~\ref{sec3} and in Ref.~\cite{Davydova2022}, where maximum efficiency is reached for a fully
transparent junction with ${\cal T}=1$.  

Finally, we note that we have also obtained preliminary results for the MAR-induced current-voltage curve for the above
model. We again find large rectification efficiencies $\eta(V)$ in the subgap regime $e|V|\alt 2\Delta$.  A detailed account will be given elsewhere.

\bibliography{sup}
\end{document}